\documentclass[a4paper,preprint]{emulateapj}
\usepackage{amstext}
\usepackage{amsmath}

\shorttitle{Afterglows Shed New Light on Nature of XRFs}
\shortauthors{Granot, Ramirez-Ruiz \& Perna}

\begin{document}

\title{Afterglow Observations Shed New Light on the Nature of X-ray Flashes}

\author{Jonathan Granot\altaffilmark{1},
Enrico Ramirez-Ruiz\altaffilmark{2,}\altaffilmark{3} and Rosalba
Perna\altaffilmark{4}}

\altaffiltext{1}{Kavli Institute for Particle Astrophysics and
  Cosmology, Stanford University, P.O. Box 20450, MS 29, Stanford, CA
  94309; granot@slac.stanford.edu} \altaffiltext{2}{Institute for
  Advanced Study, Einstein Drive, Princeton, NJ 08540; enrico@ias.edu}
\altaffiltext{3}{Chandra Fellow} \altaffiltext{4}{JILA and Department
  of Astrophysical and Planetary Sciences, University of Colorado at
  Boulder, 440 UCB, Boulder, CO 80309}

\begin{abstract}
  
  X-ray flashes (XRFs) and X-ray rich gamma-ray bursts (XRGRBs) share
  many observational characteristics with long duration ($\gtrsim
  2\;$s) GRBs, but the reason for which the spectral energy
  distribution of their prompt emission peaks at lower photon
  energies, $E_p$, is still a subject of debate.  Although many
  different models have been invoked in order to explain the lower
  values of $E_p$, their implications for the afterglow emission were
  not considered in most cases, mainly because observations of XRF
  afterglows have become available only recently.  Here we examine the
  predictions of the various XRF models for the afterglow emission,
  and test them against the observations of XRF 030723 and XRGRB
  041006, the events with the best monitored afterglow light curves in
  their respective class. We show that most existing XRF models are
  hard to reconcile with the observed afterglow light curves, which
  are very flat at early times.  Such light curves are, however,
  naturally produced by a roughly uniform jet with relatively sharp
  edges that is viewed off-axis (i.e. from outside of the jet
  aperture).  This type of model self consistently accommodates both
  the observed prompt emission and the afterglow light curves of XRGRB
  041006 and XRF 030723, implying viewing angles $\theta_{\rm obs}$
  from the jet axis of $(\theta_{\rm obs}-\theta_0)\sim 0.15\theta_0$
  and $(\theta_{\rm obs}-\theta_0)\sim \theta_0$, respectively, where
  $\theta_0\sim 3^\circ$ is the half-opening angle of the jet. This
  suggests that GRBs, XRGRBs and XRFs are intrinsically similar
  relativistic jets viewed from different angles.  It is then natural
  to identify GRBs with $\gamma(\theta_{\rm obs}-\theta_0)\lesssim 1$,
  XRGRBs with $1\lesssim\gamma(\theta_{\rm obs}-\theta_0)\lesssim{\rm
    a\ few}$, and XRFs with $\gamma(\theta_{\rm
    obs}-\theta_0)\gtrsim{\rm a\ few}$, where $\gamma$ is the Lorentz
  factor of the outflow near the edge of the jet from which most of
  the observed prompt emission arises. Future observations with {\it
    Swift} could help test this unification scheme in which GRBs,
  XRGRBs and XRFs share the same basic physics and differ only by
  their orientation relative to our line of sight.

\end{abstract}


\section{Introduction}
\label{intro}

X-ray flashes (XRFs) are transient X-ray sources with durations
ranging from several seconds to a few minutes and their distribution
on the sky is consistent with it being isotropic
\citep{Heise01,Kippen03}, similar to what is observed in long duration
($\gtrsim 2\;$sec) gamma-ray bursts (GRBs). XRFs are also similarly
variable. They were first detected by the wide field camera (WFC) of
{\it BeppoSAX} \citep{Heise01}, and subsequently studied with {\it
  HETE-II} \citep{Barraud03,Lamb04}. In addition to XRFs, {\it
  HETE-II} expanded the empirical classification of variable X-ray
transients to include an intermediate class of events known as X-ray
rich GRBs (XRGRBs). The spectrum of XRGRBs and XRFs is similar to that
of GRBs \citep{Sakamoto04} except for the lower values of the photon
energy $E_p$ at which their $\nu F_\nu$ spectrum peaks, and the lower
energy output in gamma-rays and/or X-rays, $E_{\rm\gamma,iso}$,
assuming isotropic emission. In all other respects XRFs, XRGRBs and
GRBs seem to form a continuum.

Many different models have been proposed for XRFs, most of which try
to incorporate them in a unified scenario with GRBs. These models
include high redshift GRBs \citep{Heise01}, dirty (low $\gamma$)
fireballs \citep{DCB99,Heise01,Huang02,ZWM03,ZWH04}, regular GRBs
viewed off-axis \citep{YIN02,YIN03,YIN04a,YIN04b,DDD04,CK04,ZWH04}, photosphere
dominated emission \citep{Drenkhahn02,RRLR02,Meszaros02}, week
internal shocks \citep[low variability,
$\Delta\gamma\ll\gamma$;][]{ZM02a,Barraud03,moch03}, and large viewing
angles in a structured \citep{Lamb05} or quasi \citep{Zhang04}
universal jet.

Most of these models mainly aim at explaining the low values of $E_p$
in XRFs, and do not address their expected afterglow properties. The
afterglow evolution alone can, however, serve as a powerful test for
XRF models, especially after the recent discovery of several
afterglows of XRFs (020427, 020903, 030723, 040701, 040825B, 040912,
040916) and XRGRB 041006.  Until a few years ago, XRFs were known
predominantly as bursts of X-rays, largely devoid of any observable
traces at any other wavelengths. However, a striking development in
the last several years through the impetus of the {\it HETE-II}
satellite, has been the measurement and localization of fading X-ray
and optical signals from some XRFs. These afterglow observations
resulted in three redshift determinations, for XRF 020903
\citep[$z=0.251$,][]{Soderberg04}, XRF 040701
\citep[$z=0.2146$,][]{Kelson04} and XRGRB 041006
\citep[$z=0.716$,][]{Fugazza04,Price04b}.  In two cases, XRF 030723
and XRGRB 041006, the afterglow light curves are reasonably well
monitored from sufficiently early times so that they can be used to
derive meaningful constraints on XRF models.

In this paper we critically examine the different XRF models and
contrast them with the afterglow observations of XRF 030723 and XRGRB
041006, as well as other available observations such as the prompt
emission characteristics and the measured distances.  The paper is
organized as follows. \S \ref{clas} describes the current empirical
classification and general properties of GRB, XRGRB and XRF sources.
Various XRF models are considered in \S \ref{others} along with a
brief discussion of the observations that support or undermine these
schemes. All the models that are discussed in \S \ref{others} have at
least one major flaw in common: they do not naturally produce the very
flat afterglow light curve seen at early times in both XRF 030723 and
XRGRB 041006.  In the remainder of the paper we thus concentrate only
on the class of models which naturally produce such light curves. That
is, a roughly uniform jet with sufficiently sharp edges viewed outside
the jet core. This class of models is discussed qualitatively in \S
\ref{off-axis}, and more quantitatively in \S \ref{obs}, where it is
also directly compared to the prompt emission and afterglow
observations of XRGRB 041006 (\S \ref{XRGRB041006}) and XRF 030723 (\S
\ref{XRF030723}).  The role of our viewing angle as an essential
parameter is given particular attention.  In \S \ref{other_obs} we
briefly consider other XRFs and XRGRBs, and find that the data in
these cases are too sparse and insufficient in order to derive
meaningful constraints on the underlying model.  Our conclusions are
discussed in \S \ref{conc}.

\section{Empirical Classification of GRBs, XRGRBs \& XRFs} 
\label{clas}

The operational definition of an XRF by the {\it BeppoSAX} team was
that of a transient source, with a duration of less than $10^3\;$sec,
whose flux triggered the Wide Field Camera (WFC) but not the Gamma Ray
Burst Monitor (GRBM). Later, with {\it HETE-II}, the definition
changed slightly and was based on the ratio of the fluence in the
X-ray band to that in the gamma-ray band, $f_{\rm
  X/\gamma}=\log_{10}[S_{\rm X}(2-30\;{\rm keV})/S_\gamma(30-400\;{\rm
  keV})]$. In addition to XRFs, an intermediate class of X-ray rich
GRBs (XRGRBs) was also introduced. According to this new empirical
scheme, GRBs, XRGRBs, and XRFs correspond to $f_{\rm X/\gamma}<-0.5$,
$-0.5<f_{\rm X/\gamma}<0$, and $f_{\rm X/\gamma}>0$, respectively.
Although the observed peak energies, $E_p^{\rm obs}=(1+z)^{-1}E_p$
(the photon energy where $\nu F_\nu$ peaks), are on average about a
factor of $\sim 10$ less than those of the ``standard'' GRBs
($E_{p,{\rm XRF}}^{\rm obs}\sim 25$ keV while $E_{p,{\rm GRB}}^{\rm
  obs}\sim 250$ keV), the spectra of XRFs are fitted by the same Band
function that is commonly used to fit GRBs \citep{Band93}, and they
seem to obey the same correlation between $E_p$ (that is corrected for
cosmological redshift) and the isotropic energy output seen in
gamma-rays (or X-rays), $E_{\rm\gamma,iso}$: $E_p\propto
E_{\rm\gamma,iso}^{1/2}$ \citep{Amati02,L-RR-R02,Lamb05}.  While GRBs
and XRFs have a different operational definition, they appear to form
a continuum of events, rather than a bimodal distribution, with bursts
varying uniformly from XRFs to XRGRBs to GRBs.

\section{The Viability of Various XRF Models} 
\label{others}

In the next section we show that in order to reproduce the observed
behavior seen in the afterglow light curves of both XRGRB 041006 and
XRF 030723 a roughly uniform jet with sufficiently sharp edges viewed
off-axis is required. This is a direct consequence of the very flat
evolution of the afterglow light curve that is seen at early times.
Such a behavior does not occur for a spherically symmetric outflow, or
for a uniform jet that is viewed from within its aperture. The same
also applies for a ``structured'' jet \citep{LPP01,RLR02,ZM02b} where
the energy per solid angle $\epsilon$ props as the inverse square of
the angle $\theta$ from the jet axis, outside of some small core angle
$\theta_c$, $\epsilon\approx\epsilon_0\min[1,(\theta/\theta_c)^{-2}]$.
For most models that have been proposed in the literature to explain
the phenomenology of XRFs, the afterglow light curve at early times is
expected to be similar to that of a spherical flow [with
$\epsilon=\epsilon(\theta_{\rm obs})$ if $\epsilon$ varies with
$\theta$], and thus behave qualitatively similar to GRB afterglow
light curves.  Although this early afterglow behavior alone makes most
XRF models inconsistent with observations, in what follows, we give
additional arguments that further undermine these various schemes.

A straightforward interpretation of the low $E_p^{\rm obs}$ seen in both XRFs
and XRGRBs is that they are in fact the high-redshift counterparts of
long duration GRBs.  While XRFs have on average lower energies than
GRBs, their durations are comparable to those of GRBs \citep{Heise01},
which argues against a high-redshift origin. Moreover, the recent
redshift determination of XRF 020903
\citep[$z=0.251$;][]{Soderberg04}, XRF 040701
\citep[$z=0.2146$,][]{Kelson04} and XRGRB 041006
\citep[$z=0.716$;][]{Fugazza04,Price04b} directly rules out this
interpretation. Although some GRBs at very high redshifts may resemble
XRFs, it is now clear that they do not represent the bulk of the
population. In fact, recent estimates suggest that this high-redshift
population may only constitute a small fraction of the total number of
bursts, provided that the redshift distribution of GRBs accurately
tracks the cosmic star formation rate of massive stars
\citep[e.g.,][]{BN00,BL02,LRFRR02}.

\cite{DCB99} have pointed out that ``dirty fireballs'', i.e.
relativistic outflows with a larger baryonic load and hence a lower
initial Lorentz factor $\Gamma_0$ compared to classical GRBs, would
have a smaller $E_p$ which could be in the X-rays. When XRFs were
discovered it was natural to suggest this scenario as a possible way
of achieving low values for $E_p$ \citep{Heise01,Huang02}.  We note
here that while a lower $\Gamma_0$ implies a lower $E_p$
($\propto\Gamma_0^4$) in the external shock model for the prompt
emission, in the internal shocks model it would produce a higher $E_p$
($\propto\Gamma_0^{-2}$). For the external shock model, the lower the
observed $E_p=(1+z)E_p^{\rm obs}$ is, the lower the value of
$\Gamma_0$ that is required to explain it. This implies that events
with lower values of $E_p$ should have longer durations, since the
deceleration time scales as $t_{\rm dec}\propto
\epsilon^{1/(3-k)}\Gamma_0^{-2(4-k)/(3-k)}$, where $\epsilon$ is the
energy per solid angle and $\rho_{\rm ext} \propto r^{-k}$. This is
inconsistent with observations, as no clear trend exists between the
total duration of the event and its $E_p$ \citep{Sakamoto04}. What is
more, as is the case in any external shock model, the afterglow should
be a smooth continuation of the prompt emission as both arise from the
same external shock.  The observations of XRF 030723 \citep{Fynbo04a}
offer the best evidence so far against this.

The presence of a dominant baryonic or shock pair photosphere within
the standard fireball model was invoked by \citet{RRLR02} and
\citet{Meszaros02} to explain the formation of XRFs. While this is a
tenable scenario for producing XRGRBs, the very low $E_p<5\;$keV
observed for XRFs at $z\sim 0.2$ are hard to reconcile with a
low-$\Gamma_0$, pair-dominated photospheric component
\citep{Meszaros02}.

Another way of obtaining low values of $E_p$ is with a roughly
constant $\Gamma_0$ between different events, but with a small
contrast in the value of $\Gamma_0$ between different colliding shells
in the internal shocks model, $\Delta\Gamma_0\ll\Gamma_0$
\citep{ZM02a,Barraud03,moch03}. This model, however, should produce an
afterglow with an intensity that is comparable to those seen in
typical GRBs. Furthermore, the isotropic equivalent kinetic energy of
the afterglow shock at early times (when only the local value of
$\epsilon$ along the line of sight is sampled, similar to the prompt
emission) should be much larger than $E_{\rm\gamma,iso}$, because of
the low radiative efficiency of the prompt emission in this scenario.

An alternative model for XRFs arises in the context of the so called
universal (structured) jet models.  In this class of models it is
assumed that all GRB jets have the same structure, where both
$\epsilon$, and $\Gamma$ depend on the angle $\theta$ with respect to
the jet axis \citep{LPP01,RLR02,ZM02b}.  This model can reproduce the
key features expected from the conventional on-axis uniform jet
models, with the novelty being that the achromatic break time in the
broadband afterglow light curves corresponds to the epoch during which
the core of the jet becomes visible, rather than the edge of the jet
as in the uniform jet model.  For the internal shock model, which is
thought to be the mechanism responsible for the prompt emission, $E_p
\propto L_{\rm iso}^{1/2}\Gamma_0^{-2}$ \citep[e.g.,][]{RRLR02}. As
there is no observed correlation between the duration on an event and
its $E_p$, this suggests that $E_p \propto
E_{\rm\gamma,iso}^{1/2}\Gamma_0^{-2}$.  This reproduces the observed
narrow correlation $E_p \propto E_{\rm\gamma,iso}^{1/2}$ only if
$\Gamma_0$ is both independent of $\theta$ and has a very small
scatter between different events.

\cite{Lamb05} have proposed a unified description of XRFs, XRGRBs and
GRBs in which either (i) the half-opening angle $\theta_0$ of a
uniform jet varies over a wide range while its energy remains
constant, or (ii) our viewing angle $\theta_{\rm obs}$ with respect to
a universal structured jet varies over a wide range. For convenience,
we shall refer to $\theta_0$ and $\theta_{\rm obs}$ in these two
options, , respectively, simply as $\theta_*$.  In this picture, small
values of $\theta_*$ correspond to GRBs, while increasingly larger
values of $\theta_*$ correspond to XRGRBs and then to XRFs. In this
scenario $E_p\propto\theta_*^{-1}$, so that the large range of
observed $E_p$ values, ranging from $E_p\gtrsim 1\;$MeV for bright
GRBs to $E_p\lesssim 5\;$keV for dim XRFs (i.e. a range of a factor of
$\gtrsim 200$), directly corresponds to a similar range in $\theta_*$.
Both the inferred values of $\theta_*$ from the jet break times in the
afterglow light curves, and the $\log N-\log S$ distribution of {\it
BATSE} GRBs \citep{GGB05} suggest, however, a smaller range for
$\theta_*$, of about $\sim 10$ ($0.05\lesssim\theta_*\lesssim 0.5$),
rather than $\gtrsim 200$.

\section{Off-Axis Jet Models of GRBs \& XRFs} 
\label{off-axis}

The possibility that GRB outflows are collimated into narrow jets,
where in many cases our line of sight would be outside of the jet
aperture, resulting in no detectable prompt emission and an ``orphan
afterglows'' at later times, was suggested by \citet{Rhoads97}. This
was shortly after the first detection of a GRB afterglow and before
there was compelling observational evidence for jets in GRBs.  As
observational evidence in favor of GRB outflows being collimated into
narrow jets gradually accumulated, studies of the observational
signatures of off-axis GRB jets became more common
\citep{PL98,WL99,MSB00,DGP02,Granot02,Levinson02,TP02,NPG03,GL03,EL04},
again mostly devoted to orphan afterglows. The possibility that for
viewing angles that are only slightly outside of the jet aperture the
prompt emission might still be detectable, but would shift into the
X-rays due to the reduced Doppler factor, has been pointed out by
\citet{WL99}. That was, however, before the discovery of XRFs. After
XRFs were discovered, \citet{YIN02} suggested that GRB jets viewed
slightly off-axis could naturally account for this newly discovered
class of events.  In later works they have significantly developed
some aspects of this model \citep{YIN03,YIN04a,YIN04b}, in particular
those regarding the prompt emission. In this section we discuss
various aspects of this model in some detail, with both the prompt and
afterglow signatures being at the forefront of our attention.

\subsection{The Jet Structure}
\label{jet_structure}

The usual assumption about the jet structure is that it is perfectly
uniform within some finite initial half-opening angle, $\theta_0$,
from the jet symmetry axis, and abruptly truncates outside of
$\theta_0$ \citep{WL99,YIN02}. Obviously, this is only an
approximation, as physically one might expect that the jet would have
a smoother outer edge, where the energy per solid angle, $\epsilon$,
and the initial Lorentz factor, $\Gamma_0$, decrease smoothly with the
angle $\theta$ from the jet symmetry axis, over some finite range in
$\theta$, $\Delta\theta\gtrsim\Gamma_0^{-1}$. In fact, numerical
simulations show that even if the jet initially has perfectly sharp
edges (i.e. a `top hat' jet), the interaction with the ambient medium
causes its edges to become smoother with time \citep{Granot01}. This
serves as a motivation for considering a roughly uniform jet with
smooth edges, as a more realistic version of the `top hat' jet. The
most widely used version of such a jet is one with a Gaussian angular
profile for $\epsilon$ \citep{ZM02b,KG03},
$\epsilon=\epsilon_0\exp(\theta^2/2\theta_0^2)$.  There are also other
similar jet profiles, where most of the energy resides within some
finite half-opening angle $\theta_0$, and $\epsilon$ sharply drops
outside of $\theta_0$. Numerical simulation of a jet boring its way
through a massive star progenitor in the context of the collapsar
model \citep{ZWH04} predict a roughly uniform jet core with
$\theta_0\sim 3^\circ-5^\circ$ and wings where
$\epsilon\propto\theta^{-3}$ which extend to larger angles, i.e.
$\epsilon\approx\epsilon_0\min[1,(\theta/\theta_0)^{-3}]$. We consider
all such models to be variants of the same basic jet structure, and
when viewed from outside the jet core ($\theta_{\rm obs}>\theta_0$)
they are considered as members of the same class of XRF models. The
different aforementioned variants of this jet structure are considered
in \S \ref{obs}.

\subsection{The Afterglow Light Curves}
\label{AG}

\begin{figure}
\plotone{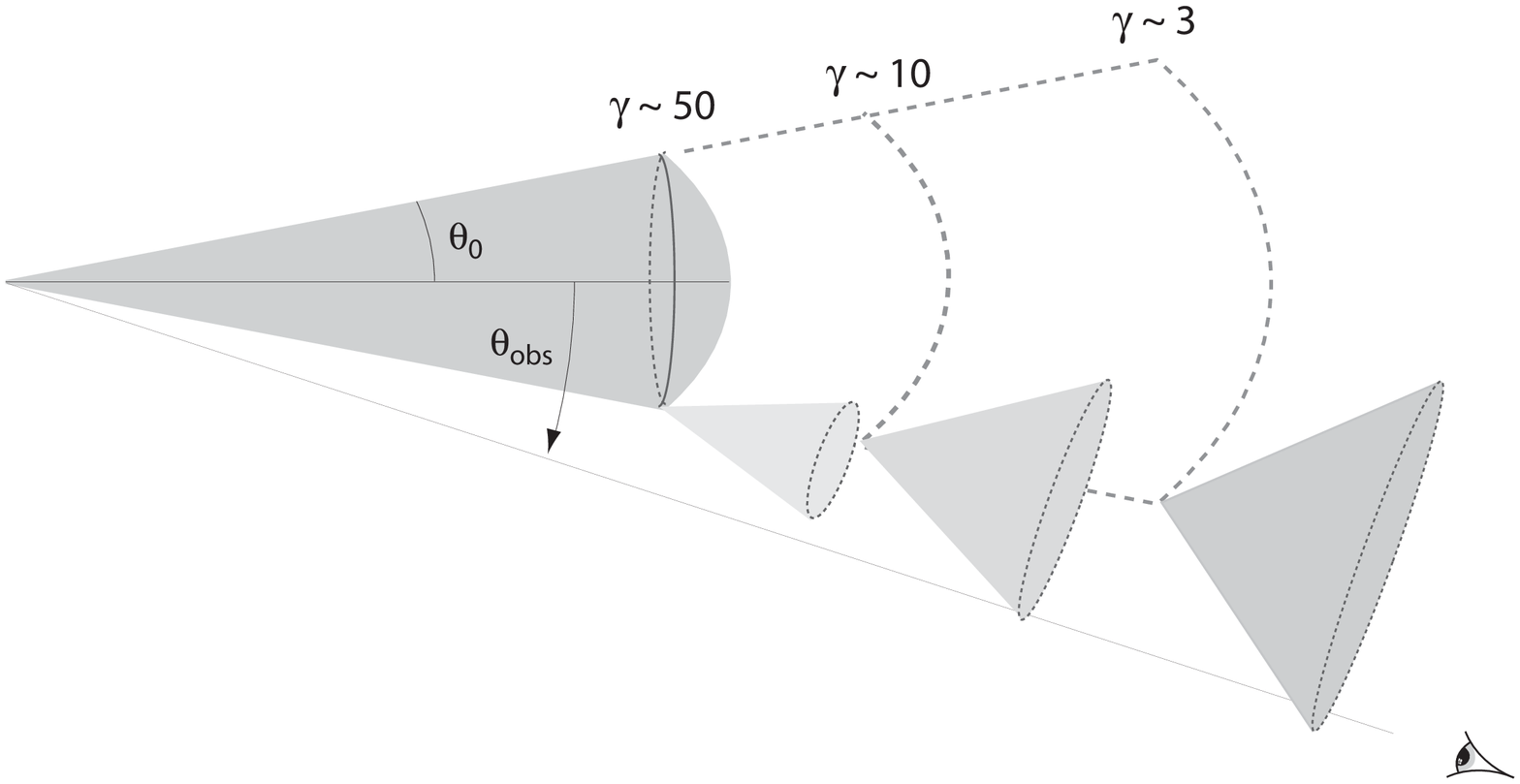}
\caption{\label{diagram} An illustrative diagram of the emission from
  a uniform relativistic jet with sharp edges and half opening angle
  $\theta_0$, that is seen by an off-axis observer whose line of sight
  makes an angle $\theta_{\rm obs}>\theta_0$ with the jet
  axis. Because of relativistic beaming (i.e. aberration of light) the
  emission from each part of the jet is beamed into a narrow cone of
  half-opening angle $\gamma^{-1}$ around its direction of motion in
  the observer frame. During the prompt emission (and the very early
  afterglow) the Lorentz factor of the jet is large ($\gamma\gtrsim
  50$) and therefore most of the radiation is strongly beamed away
  from the line of sight. In this case, the little radiation that is
  observed comes mainly from near the edge of the jet, at the point
  closest to the line of sight. As the jet decelerates $\gamma$
  decreases with time and the beaming cone grows progressively wider,
  causing the radiation to be less strongly beamed, resulting in a
  rising light curve. The light curve peaks when $\gamma$ drops to
  $\sim(\theta_{\rm obs}-\theta_0)^{-1}$ as the line of sight enters
  the beaming cone of the emitting material at the edge of the jet
  (the middle beaming cone in the figure), and subsequently decays
  with time, asymptotically approaching the light curve for an on-axis
  observer ($\theta_{\rm obs}<\theta_0$) at later times.}
\end{figure}

The early afterglow light curves for off-axis viewing angles
($\theta_{\rm obs}>\theta_0$) are generally flatter than those
observed in typical on-axis ($\theta_{\rm obs}<\theta_0$) GRB
afterglows \citep[see][and references therein]{Granot02}. For a jet
structure for which $\epsilon$ and $\Gamma_0$ drop sharply with
$\theta$ at $\theta>\theta_0$, we expect its early light curve to rise
with time. In this case, the sharper the edge of the jet, the sharper
the rise in the light curve \citep{Granot02}. For jets with sharp
enough edges, the emission from the core of the jet (i.e. from
$\theta<\theta_0$) dominates even at off-axis viewing angles
($\theta_{\rm obs}>\theta_0$), despite it being strongly beamed away
from our line of sight (see Fig.  \ref{diagram}). This is either
because there is no emitting material along the line of sight, or even
if present its emission is still weaker than that arising from the jet
core. As the jet sweeps up an increasing amount of external medium, it
slows down and thereafter the relativistic beaming of the emission
from the jet core away from our line of sight decreases. When $\gamma$
drops to $\sim (\theta_{\rm obs}-\theta_0)^{-1}$, our line of sight
enters the beaming cone of the radiation from the jet core, causing
the light curve to peak and subsequently decay, asymptotically
approaching the light curve for an on-axis observer.

If the edge of the jet is not sufficiently sharp (i.e. if $\epsilon$
and $\Gamma_0$ do not drop sufficiently sharply with $\theta$ at
$\theta>\theta_0$), then the emission from material along our line of
sight may dominate over that from the core of the jet for viewing
angles slightly outside the edge of the jet. In this case the light
curve at early times would not rise with time, but would instead
simply decay more slowly when compared to the light curve seen by
on-axis observers ($\theta_{\rm obs}<\theta_0$).  Therefore, we
conclude that the jet structure, and specifically the sharpness of its
edges, can be constrained by early afterglow observations. In the
context of the model discussed in this section, increasingly larger
viewing angles will correspond to XRGRBs and XRFs. Such a scheme is
tested against observations of XRGRB 041006 and XRF 030723 in \S
\ref{obs}.

\subsection{The Prompt \& Reverse Shock Emission}
\label{prompt}

The prompt emission for off-axis viewing angles ($\theta_{\rm
  obs}>\theta_0$) may also be dominated either by the emission from
  the jet core or by the emission from the material along the line of
  sight, depending on the viewing angle and on how sharp the edge of
  the jet is. If the edge of the jet is sufficiently sharp, the prompt
  emission is dominated by the core of the jet, and both the fluence
  and the peak photon energy drop sharply when compared to their
  on-axis values, as $[\gamma(\theta_{\rm obs}-\theta_0)]^{-6}$ and
  $[\gamma(\theta_{\rm obs}-\theta_0)]^{-2}$, respectively
  \citep{Granot02,R-R05}. The prompt emission in this case arises from
  the same region as for on-axis viewing angles, which in this
  scenario correspond to GRBs.  This suggests that the same physical
  mechanism is responsible for the prompt emission in GRBs and in XRFs
  (i.e. most likely internal shocks).

If, on the other hand, the edges of the jet are not sharp enough, then
the prompt emission will be dominated by material along our line of
sight. As it might be hard to produce strong variability in the
Lorentz factor of the outflow outside the core of the
jet \citep{RR02,ZWH04}, internal shocks may not be very efficient, and
the external shock due to the interaction with the external medium
might dominate the prompt emission. In that case, a smooth prompt
light curve consisting of a single wide peak might be expected.

The `optical flash' emission from the reverse shock is generally
expected be weaker for off-axis observers \citep{FWW04}. If the
reverse shock is Newtonian or only mildly relativistic, then the
beaming of the prompt emission (that is attributed to internal shocks
within the outflow, which occur before the ejecta is decelerated by
the external medium) and the reverse shock emission would not be very
different. In this case the ratio of the off-axis to on-axis flux or
fluence should be roughly similar for the optical flash and the prompt
emission.  If the reverse shock is relativistic then it would
significantly decelerate the ejecta, and the emission from the reverse
shock would be less strongly beamed than the prompt emission. In this
case, if the emission is dominated by the jet core (i.e.  for a sharp
edged jet), the `optical flash' emission at off-axis viewing angles
could be less suppressed compared to the prompt X-ray or gamma-ray
emission.

\subsection{Linear Polarization}
\label{pol}

An interesting implication of the off-axis jet model for XRGRBs and
XRFs is that it predicts a higher degree of linear polarization of the
prompt emission, the emission from the reverse shock, and the
afterglow emission, if the polarization is dominated by the jet
geometry while the magnetic field is mostly tangled in the plane of
the shock, as expected from the two stream instability \citep{ML99}.
For a shock produced magnetic field that is tangled within the shock
plane, the polarization peaks at a viewing angle that satisfies
$\gamma(\theta_{\rm obs}-\theta_0)\sim 1$
\citep{Gruzinov99,Waxman03,Granot03,NPW03}, since at such a viewing
angle most of the observed radiation is emitted roughly along the
shock plane in the rest frame of the emitting plasma, due to
aberration of light effects. The peak polarization can reach up to
tens of percent. This is relevant to the prompt emission, and may also
be relevant for the optical flash emission. The peak of the
polarization which occurs at $\gamma(\theta_{\rm obs}-\theta_0)\sim 1$
can shift to a larger viewing angle $\theta_{\rm obs}$ during the
optical flash as the ejecta is decelerated by the reverse shock.

For jets with sufficiently sharp edges that are viewed off-axis, the
afterglow light curve initially rises at early times, and the
polarization peaks around the time of the peak in the light curve,
which occurs when $\gamma(\theta_{\rm obs}-\theta_0)$ decreases to
$\sim 1$, as our line of sight enters the beaming cone of the emitting
material \citep{Granot02}.  Even if there is some lateral spreading of
the jet, and an initially off-axis viewing angle enters into the jet
aperture as the latter grows with time, then the afterglow
polarization would be relatively large, as the line of sight would
still be relatively close to the edge of the jet \citep{Sari99,GL99}.

One should keep in mind, however, that ordered magnetic fields might
potentially play an important role in the polarization of the prompt
emission \citep{Waxman03,Granot03,NPW03,LPB03} as well as that of the
reverse shock emission (the `optical flash' and 'radio flare') and the
afterglow emission \citep{GKo03}. If the dominant cause of polarization
is an ordered magnetic field component, instead of the jet geometry
together with a shock produced magnetic field, then the viewing angle
would have a smaller effect on the observed linear polarization.

A recent analysis of archival `radio flare' observations \citep{GT05}
has set strong upper limits of the linear and circular polarization of
this radio emission, and showed that these limits constrain the
presence of an ordered magnetic field in the ejecta. The existing
radio flare observations are for GRBs, which in the model considered
here correspond to on-axis viewing angles ($\theta_{\rm
obs}<\theta_0$).  For a uniform jet with an ordered toroidal magnetic
field, the polarization vanishes at the jet symmetry axis (i.e. for
$\theta_{\rm obs}=0$) and strongly increases toward the edge of the
jet.  Therefore, the observed upper limits on the linear polarization
translate to an upper limit on $\theta_{\rm obs}/\theta_0$. The best
constraints so far are for GRB 991216: $P<7\%$ and $\theta_{\rm
obs}/\theta_0\lesssim 0.4-0.55$, respectively. There are weaker
constraints for GRBs 990123 and 020405. Tighter constraints on the
presence of an ordered magnetic field in the ejecta are expected in
the near future when a larger sample of radio flare polarization
measurements becomes available.  Such measurements for XRGRBs or XRFs
are crucial when testing the off-axis jet model. This is because in
this model one expects a viewing angle that is only slightly outside
the edge of the jet and thus a large degree of polarization (tens of
percent) for a purely ordered toroidal magnetic field in the ejecta.

\subsection{Description of the Numerical Model}
\label{num}

In this section we briefly describe the model that is used in \S
\ref{obs} for describing the data. This is essentially model 1 of
\citet{GK03}, similar to that used by \citet{R-R05} for modeling the
lightcurve of GRB 031203.  The deceleration of the flow is calculated
from the mass and energy conservation equations and the energy per
solid angle $\epsilon$ is taken to be independent of time. The local
emissivity is calculated using the conventional assumptions of
synchrotron emission from relativistic electrons that are accelerated
behind the shock into a power-law distribution of energies,
$N(\gamma_e) \propto \gamma_e^{-p}$ for $\gamma_e > \gamma_m$, where
the electrons and the magnetic field hold fractions $\epsilon_e$ and
$\epsilon_B$, respectively, of the internal energy. The external
density is taken to be a power law in the distance $r$ from the
central source, $\rho_{\rm ext}=Ar^{-k}$, where $k=0$ corresponds to a
uniform ISM while $k=2$ corresponds to a stellar wind of a massive
star progenitor (assuming a constant ratio for the mass loss rate and
the wind velocity). Another important physical parameter is the (true)
energy of the jet, $E$, which is calculated assuming that the jet is
double sided.  The synchrotron spectrum is taken to be a piecewise
power law \citep{SPN98}.  The inverse-Compton scattering of the
synchrotron photons by the same relativistic electrons, that is known
as synchrotron-self Compton (SSC), is also taken into account
\citep{PK00,SE01}.

The lateral spreading of the jet is neglected in this model.  This
approximation is consistent with results of numerical studies
\citep{Granot01,KG03} which show relatively little lateral expansion
as long as the jet is relativistic. The light curves for observers
located at different angles, $\theta_{\rm obs}$, with respect to the
jet axis are calculated by applying the appropriate relativistic
transformation of the radiation field from the local rest frame of the
emitting fluid to the observer frame and integrating over equal photon
arrival time surfaces \citep{Granot02,R-RM04}.

\section{Observations}
\label{obs}

The goal of this section is to quantitatively test the idea that a
relativistic jet pointing slightly away from us could explain the
observations of XRGRBs and XRFs. The modeling of radio, optical, and
X-ray data is carried out in the framework of collimated ejecta
interacting with the external medium. The model is described in \S
\ref{off-axis}.  In this work we focus our attention on two afterglows
for which radio, optical, and X-ray light curves are available: XRGRB
041006 (\S \ref{XRGRB041006}) and XRF 030723 (\S \ref{XRF030723}).
Although, as described in \S \ref{other_obs}, afterglow emission has
also been detected for other other XRFs and XRGRBs, the data in the
these cases are too sparse and insufficient in order to derive
meaningful constraints on the underlying model.

\subsection{X-ray Rich GRB 041006}
\label{XRGRB041006}

\begin{figure}
\plotone{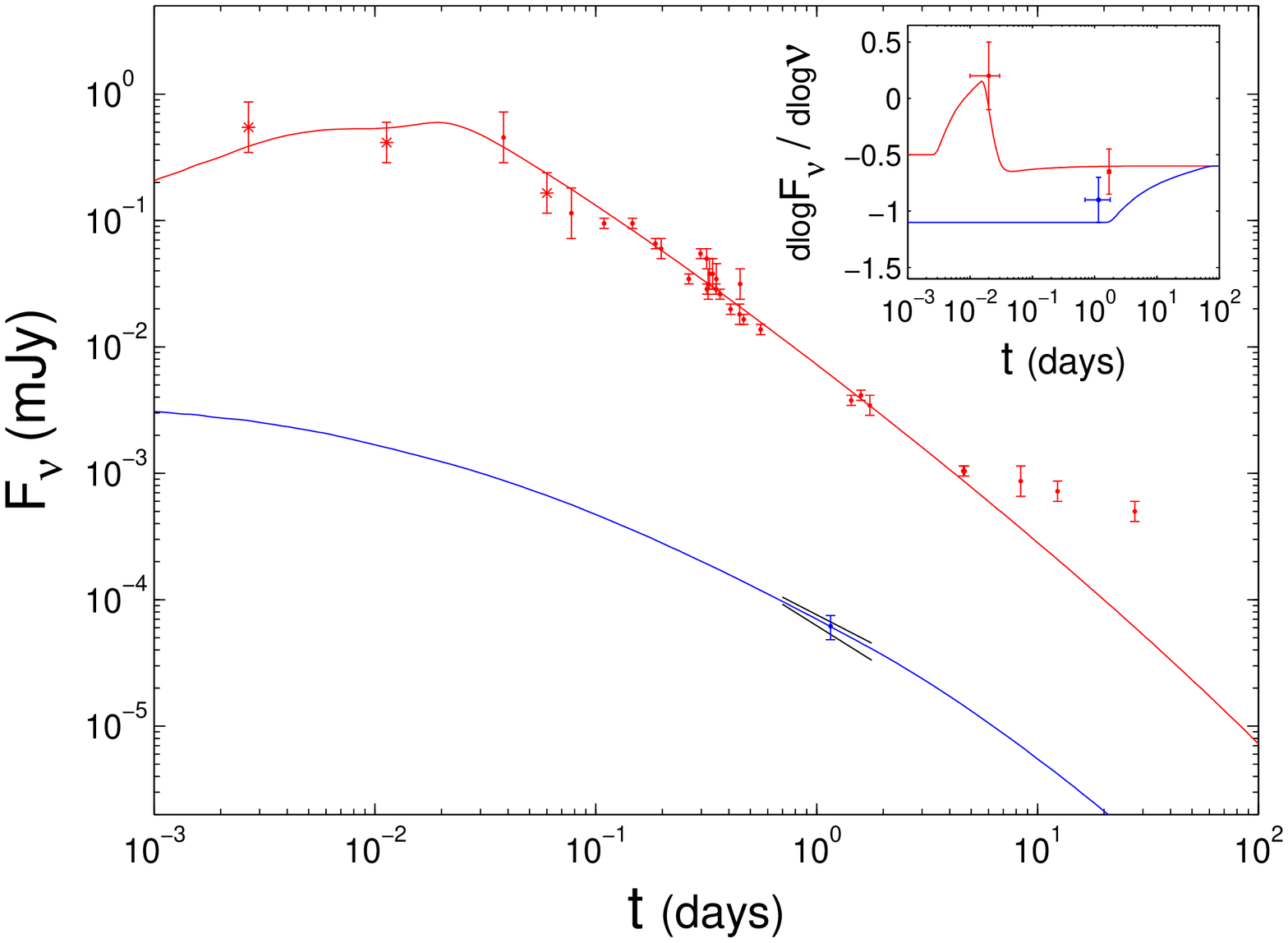}
\caption{\label{fit_XRGRB041006} A tentative fit to the optical R-band
  \citep[shown in red;][]{Ayani04,
  DaCosta04,D'Avanzo04,Ferrero04,Fynbo04b,Fugazza04,Fukushi04,Garg04,
  Greco04,Kahharov04,Kinoshita04,Klotz04,Misra04a,Misra04b,Monfardini04,
  Price04b,Yost04} and X-ray \citep[$0.5-6\;$keV, shown in
  blue;][]{Butler04b} light curves of XRGRB 041006. The ROTSE-IIIa
  points are shown with asterisk symbols since they are unfiltered,
  but they can still be treated as R-band observations within the
  measurement errors. We also added two black lines which indicate the
  edges of the $1\;\sigma$ confidence interval for the temporal decay
  index, $\alpha=1\pm 0.1$, and cover the duration of the {\it
  Chandra} observation. The inset shows the predicted spectral slope,
  $-\beta=d\log F_\nu/d\log\nu$, in the optical (red) and in the X-ray
  (blue), together with the values inferred from observations.}
\end{figure}

XRGRB 041006 was detected by {\it HETE-II} \citep{Galassi04}. It had a
fluence of $5\times 10^{-6}\;{\rm erg\;cm^{-2}}$ in the $2-30\;$keV
range and $7\times 10^{-6}\;{\rm erg\;cm^{-2}}$ in the $30-400\;$keV
range, corresponding to $f_{X/\gamma}\approx -0.15$ which classifies
it as an XRGRB. It has a redshift of $z=0.716$
\citep{Fugazza04,Price04a}, which for a fluence of $f\approx 1.2
\times 10^{-5}\;{\rm erg\;cm^{-2}}$ in the $2-400\;$keV range gives
$E_{\rm\gamma,iso}\approx 1.6\times 10^{52}\;$erg. It had an observed
peak photon energy
of\footnote{http://space.mit.edu/HETE/Bursts/GRB041006/} $E_p^{\rm
  obs}=63^{+7}_{-5}\;$keV, corresponding to $E_p=109_{-9}^{+12}\;$keV.
Figure \ref{fit_XRGRB041006} shows an off-axis model yielding an
acceptable fit to the to the optical and X-ray afterglow observations
of XRGRB 041006, which is also consistent with the upper limits at
radio and sub-mm wavelengths \citep{Barnard04a, Barnard04b, SF04}.
From this analysis one can conclude that a successful model for the
afterglow of XRGRB 041006 is that of a collimated, misaligned jet
interacting with a stellar wind external medium of mass density
$\rho_{\rm ext}=Ar^{-2}$, where $r$ is the distance from the central
source.  The parameter values used in this fit are: $E=1.0\times
10^{51}\;$erg, $A_*\equiv A/(5\times 10^{11}\;{\rm
  gr\;cm^{-1}})=0.03$, $\theta_0=3^\circ$, $\theta_{\rm
  ons}=1.15\theta_0$, $p=2.2$, $\epsilon_e=0.1$, and
$\epsilon_B=0.001$.

The optical light curve is very flat at early times ($\alpha\sim 0$ at
$t\lesssim 1\;$hr, where $F_\nu\propto t^{-\alpha}\nu^{-\beta}$) and
becomes steeper after a few hours ($\alpha\approx 1.2$), which is a
little steeper than the decay index in the X-ray at a similar time
($\alpha\approx 1$ at $t\approx 1\;$day). Also, the ratio of the flux
in the optical and X-ray at $t\approx 1\;$day implies a spectral index
of $\beta\approx 0.7-0.75$ assuming a single power law between them.
This suggests that the cooling break frequency $\nu_c$ is above the
optical after $1\;$day. Since one requires very extreme parameters to
get $\nu_c$ to the X-ray range after $1\;$day (even getting $\nu_c$ to
be above the optical after a day requires relatively low values of
$\epsilon_B$ and of the external density), it is most likely that
$\nu_c$ is between the optical and X-ray at $1\;$day, which can also
explain the steeper temporal decay index in the optical (by
$\Delta\alpha=0.25$) for a stellar wind environment ($k=2$). This
favors a wind medium over a uniform density one, since otherwise the
flux in the optical will decay more slowly than in the X-ray (also by
$\Delta\alpha=0.25$), which is contrary to what is observed for XRGRB
041006. At $t\gtrsim 5\;$days there is a flattening in the optical
light curve, which is probably due to an underlying SN component
\citep{Garg04}. This explains why the observed flux is higher than
that predicted by our narrow relativistic jet model.

The fit to the afterglow observations does not, however, uniquely
determine the model parameters. Some physical parameters are
nonetheless constrained better than others. The afterglow data for
XRGRB 041006 requires a stellar wind environment ($k=2$) with a low
density ($A_*\sim 0.03$) and a viewing angle that is only slightly
outside the edge of the jet, $(\theta_{\rm obs}-\theta_0)\sim
0.15\theta_0\sim 10^{-2}\;$rad, in order to successfully explain both
the spectrum + temporal decay rates in the optical and X-ray at $\sim
1\;$day and the very flat optical light curve seen at early times.
   
\begin{figure}
\plotone{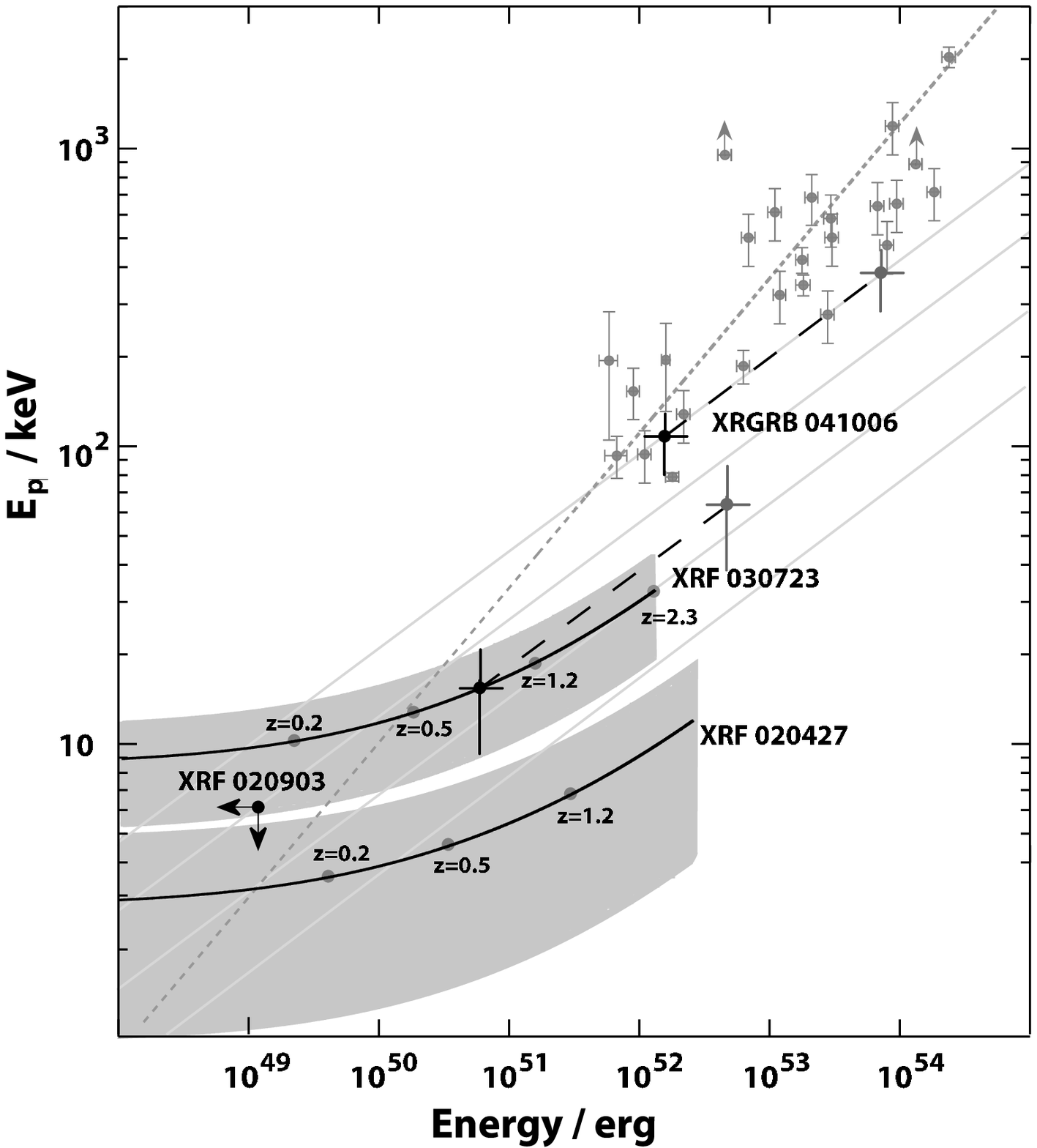}
\caption{\label{Amati_relation}XRFs in the $E_p-E_{{\gamma},{\rm
      iso}}$ plane, together with the GRBs and an X-ray rich GRBs. The
      compilation of observed $E_p$ and $E_{\gamma,{\rm iso}}$
      in the source frame derived by \citet{GGL04} are also
      illustrated.  If XRGRB 041006 was viewed on-axis (at
      $\theta_{\rm obs} < \theta_0$), the peak of the spectrum and the
      isotropic equivalent energy would be $\sim 390 \;$keV and $\sim
      7.3 \times 10^{53}\;$ergs, respectively (gray symbol). When
      viewed off-axis GRBs move in the $E_p-E_{{\gamma},{\rm iso}}$
      plane shown in the figure along straight lines (for the log-log
      axis) given by $E_p \propto E_{{\gamma},{\rm iso}}^{1/3}$
      (dashed lines).  }
\end{figure}

If GRB jets have well-defined edges, both the prompt gamma-ray fluence
and the peak of the spectrum drop very sharply outside the opening of
the jet, 
as\footnote{This is an approximate expression which
  is valid for a point source at the edge of the jet at the point
  closest to the line of sight, and gives reasonable off-axis light
  curves \citep{Granot02}.  A more accurate calculation
  \citep[e.g.,][]{EL04} shows a more complex behavior. If one defines
  the local slope of the fluence, $f$, as a function of $\delta$,
  $a=-d\log f/d\log\delta$, then $a>3$ at very small off-axis angles
  $0<\gamma(\theta_{\rm obs}-\theta_0)\lesssim 1$, $a\approx 2$ at
  intermediate angles $\gamma^{-1}<(\theta_{\rm
    obs}-\theta_0)<\theta_0$, and $a\approx 3$ at $(\theta_{\rm
    obs}-\theta_0)\gtrsim \theta_0$. This is somewhat different from
  our simple power law approximation. However, since exact shape of
  the edge, as well as other model uncertainties, could introduce
  effects of similar magnitude to the difference between our simple
  power law approximation and the more accurate calculation, the
  former is sufficient for our purposes.} $\delta^{-3}$ and
$\delta^{-1}$, respectively, where\footnote{This is the ratio of the
  Doppler factor for a viewing angle along the edge of the jet (i.e.
  at the point where most of the off-axis emission comes from; the
  Lorentz factor $\gamma$ is that of the emitting fluid at the edge of
  the jet), $\gamma|\theta_{\rm obs}-\theta_0|\lesssim 1$, and the
  Doppler factor for an off-axis viewing angle $\theta_{\rm obs}$
  which satisfies $\gamma(\theta_{\rm obs}-\theta_0)\gtrsim 1$.}
$\delta\sim[\gamma(\theta_{\rm obs}-\theta_0)]^2$.  Therefore, the low
$E_{\rm\gamma,iso}$ of XRGRB 041006 combined with $E_{\rm k,iso} =
E/(1-\cos\theta_0) \approx E(2/\theta_0^2)\approx 7.3 \times
10^{53}\;$erg implies $\delta\sim(E_{\rm
  k,iso}/E_{\rm\gamma,iso})^{1/3}\sim 3.6$ and $\gamma\sim(E_{\rm
  k,iso}/E_{\rm\gamma,iso})^{1/6}(\theta_{\rm obs}-\theta_0)^{-1}\sim
240$. This implies a (cosmological) rest frame $E_p \sim 390\;$keV,
which falls closely within the observed $E_p-E_{\rm\gamma,iso}$
relationship reported by \citet{Amati02}, \citet{L-RR-R02} and
subsequently \citet{Lamb05} using data from {\it BeppoSAX}, {\it
  BATSE} and {\it HETE-II}, respectively. This relationship finds that
in GRBs, $E_p \propto E_{\rm\gamma,iso}^{1/2}$, although a significant
amount of outliers may be present due to selection effects
\citep{Nakar05,Band05}. Figure \ref{Amati_relation} shows the location
of GRBs, XRFs and XRGRBs in the $E_p - E_{\gamma,{\rm iso}}$ plane. To
date it has been difficult to extend this relationship into the XRF
regime (especially at very low $E_p<10$ keV) since only one XRF in
this spectral energy range that has a firmly established redshift
\citep[XRF 020903 at $z=0.251$;][]{Soderberg04}. On the other hand,
the existence of XRF 030723 and XRF 020427 with $E_p^{\rm
  obs}<10\;$keV is not sufficiently constraining, since their redshift
is not known.

\subsection{XRF 030723}
\label{XRF030723}

XRF 030723 was also detected by the {\it HETE-II} satellite. It had an
observed peak photon energy of $E_p^{\rm obs}=8.4^{+3.5}_{-3.4}\;$keV
and a fluence of $f\approx 5.7\times 10^{-7}\;{\rm erg\; cm^{-2}}$ in
the $2-400\;$keV range \citep{Butler04a}. No redshift determination
has been made, although a firm upper limit of $z<2.3$ could be placed
\citep{Fynbo04a}. {\em Chandra} observations of the X-ray afterglow
were reported by \citet{Butler04a}. In the radio band, only an upper
limit of $180\;\mu$Jy was reported at $8.46\;$GHz, $3.15\;$days after
the event \citep{SBF03}.

The optical transient was discovered by \citet{Fox03}, and extensive
follow up in the optical and near-infrared was reported by
\citet{Fynbo04a}. The well monitored R-band light curve is initially
very flat\footnote{ROTSE-III performed early unfiltered optical
observation of XRF 030723 \citep{Smith03} and conclude that ``We find
no convincing evidence for a detection of the OT in the first four of
our images, but the last two images do yield marginal possible
detections''.  Therefore, in what follows, we regard them as rough
upper limits.}, with $\alpha\sim 0$ (where $F_\nu\propto
t^{-\alpha}\nu^{-\beta}$). After about $1\;$day it steepens to
$\alpha\approx 2$. This behavior is unusual for standard GRB light
curves and allows one to constrain models of XRFs.  \citet{Fynbo04a}
already noted how the early time flattening of the light curve might
be an indication of an off-axis jet. Between $1-4\;$days the optical
spectral slope $\beta_{\rm op}$ was in the range $\sim 1.0-1.3$, which
is not unusual for GRB afterglows.

After about $\sim 10\;$days, a strong bump appeared in the optical
light curve. This was assumed to be a SN component by
\citet{Fynbo04a}, while \citet{Huang04} interpreted it as an
indication of a second jet, within the context of the two component
jet model \citep[see][and references therein]{RR02,PKG05}. The bump
had a sharp rise and red colors \citep{Fynbo04a}. The sharp rise, with
$\Delta t<t$, is hard to explain in both of these models, although
\citet{Tominaga04} were able to fit the sharp rise with models of SN
light curves. The red colors arise naturally for a SN but are very
hard to account for with a two component jet model, or for this matter
also in other models for bumps in the afterglow light curve such as a
density bump in the external medium, angular inhomogeneities in the
jet (``patchy shell''), or a refreshed shock
\citep{rm98,pmr98,rmr01,wl,enrico,lazzati2,heyl,nakar,mes98,kumar}.
Therefore, the SN explanation for the bump in the optical light curve
seems to be favored by the data.

\begin{figure}
\plotone{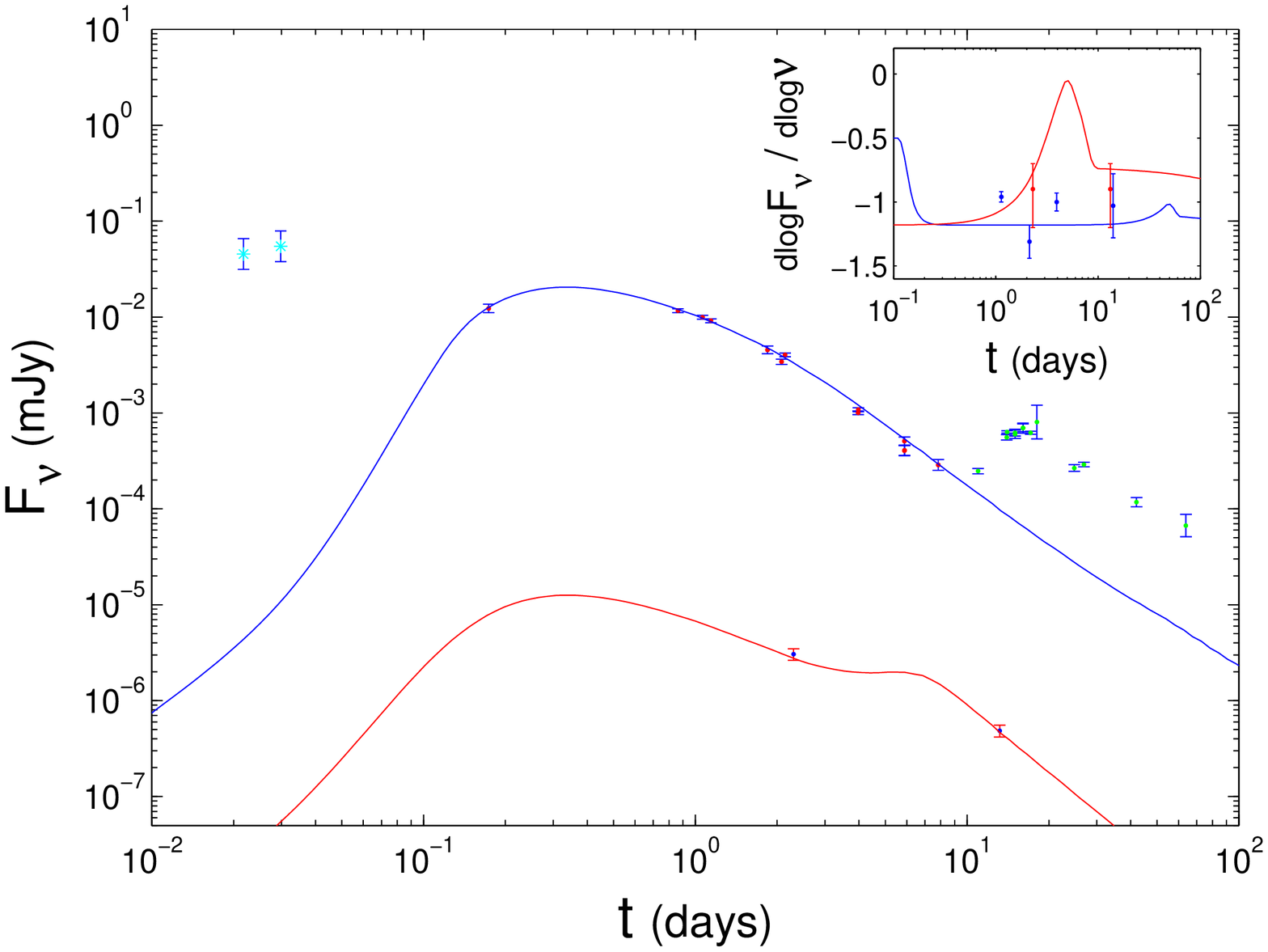} 
\caption{\label{fit XRF030723}A tentative fit
to the optical (R-band) and X-ray ($0.5-8\;$keV) light curve for XRF
030723. The first two optical points by ROTE-III are ``marginal
possible detections'' \citep{Smith03}, and are regarded as upper
limits. We do not attempt to fit the bump in the optical at $t\gtrsim
10\;$ days, as it is attributed to a separate physical component (most
likely a SN). The inset shows the spectral slope, $-\beta=d\log
F_\nu/d\log\nu$, in the optical (blue) and in the X-ray, together with
the values inferred from observations \citep{Fynbo04a,Butler04a}.}
\end{figure}

The X-ray light curve consists of two points, at $3.2\;$days and
$13.2\;$days. A joint fit for the spectral slope at these two epoch
gives $\beta_X=0.9^{+0.3}_{-0.2}$, while the temporal index between
these two points is $\alpha_X=1\pm 0.1$ \citep{Butler04a}. This is a
significantly shallower decay compared to that in the optical prior to
the bump ($\alpha_{\rm op}\approx 2$), and is therefore not easy to
account for. Since the optical bump is most likely due to a SN
component, the same physical component is not expected to contribute
significantly to the X-ray flux. The shallower decay in the X-rays
might be due to the contribution of synchrotron self-Compton (SSC)
which can dominate the X-ray flux on time scales of days to weeks
\citep{PK00,SE01}. This would generally also decrease the value of the
spectral slope, $\beta_X$, and therefore \citet{Butler04a} considered
this option to be incompatible with the data.

We performed a tentative fit to the data and demonstrate here that the
observational constraints on the spectral slope can still be satisfied
by this scenario (see Fig. \ref{fit XRF030723}).\footnote{It is also
  roughly consistent with the single upper limit in the radio, since
  the observed frequency ($8.46\;$GHz) is somewhat below the self
  absorption frequency, and scintillations may further reduce the
  observed flux.}  The physical parameters of this fit are $z=0.8$,
$E=1.0\times 10^{50}\;$erg, $n=4.5\;{\rm cm^{-3}}$, $p=2.36$,
$\epsilon_B=0.012$, $\epsilon_e=0.13$, $\theta_0=2.9^\circ$,
$\theta_{\rm obs}=2.03\theta_0$. We stress that the model parameters
cannot be uniquely determined from the fit to the afterglow
observations, and other sets of model parameters could provide an
equally good fit to the data.  Some features are, however, rather
robust. Most noticeable is a viewing angle of $\theta_{\rm obs}\sim
2\theta_0$ which is required in order to reproduce the initially very
flat part of the optical light curve. A narrow jet with $\theta_0$ of
no more than a few degrees is required in order for the jet break time
$t_j$ to be less than about a day, which is in turn needed in order to
reproduce the steep decay in the optical light curve that starts after
$\sim 1\;$day.

A redshift of $z\lesssim 0.8$ is suggested by a fit of the late time
bump in the optical light curve to core collapse SN light curves
\citep{Tominaga04}. This in part motivated us to choose a redshift of
$z=0.8$ for the fit that we present here, but fits for other values of
$z$ are also plausible.  A higher $z$ would require a higher jet
energy $E$, while a lower $z$ would require a smaller jet energy. For
$z\approx 0.8$, $E_{\rm\gamma,iso}\approx 9.3\times 10^{50}\;$erg
which together with $E_{\rm k,iso} \approx 7.8\times 10^{52}\;$erg
implies $\delta\sim 4.4$ and $\gamma\sim 40$.  This would in turn
imply a (cosmological) rest frame $E_p$ of $\sim 66\;$keV if viewed
on-axis, which is a factor of $\sim 3$ lower than the value required
to fall exactly on the Amati relation (see
Fig. \ref{Amati_relation}). Given the large uncertainties associated
with this relationship \citep{L-RR-R02,Nakar05,Band05}, we consider
this to be in good agreement with observations of on-axis GRBs.

It is reasonable to expect a relatively low Lorentz factor
($\gamma\sim 40$) at the edge of the jet. Assuming $\gamma$ decreases
from $\gamma_{\rm int}\gtrsim 100$ in the interior of the jet to much
lower values at $\Delta\theta\gtrsim 1/\gamma_{\rm int}$ centered
around $\theta_0$, then for $\gamma\lesssim 40$ the optical depth to
pair production would be large, while for larger values of $\gamma$
much fewer photons would reach and off-axis observer, so that it is
reasonable that the off-axis prompt emission will be dominated by
$\gamma$ for which $\tau_{\gamma\gamma}$ is just smaller than 1. A
similar result was obtained in a fit to GRB 031203 \citep{R-R05}.
\footnote{For XRGRB 041006 we obtain $(\theta_{\rm obs}-\theta_0)\sim
  0.15\theta_0\sim 10^{-2}\;$rad and $\gamma\sim 240$ at the edge of
  the jet. The larger value inferred for $\gamma$ might be explained
  by the smaller value of the off-axis viewing angle, $\theta_{\rm
    obs}-\theta_0$, since such a line of sight which is significantly
  closer to the edge of the jet intersects the beaming cone of the
  emitting material near the edge of the jet up to a Lorentz factor of
  $\gamma\sim(\theta_{\rm obs}-\theta_0)^{-1}\sim 10^2$.}

\begin{figure}
\plotone{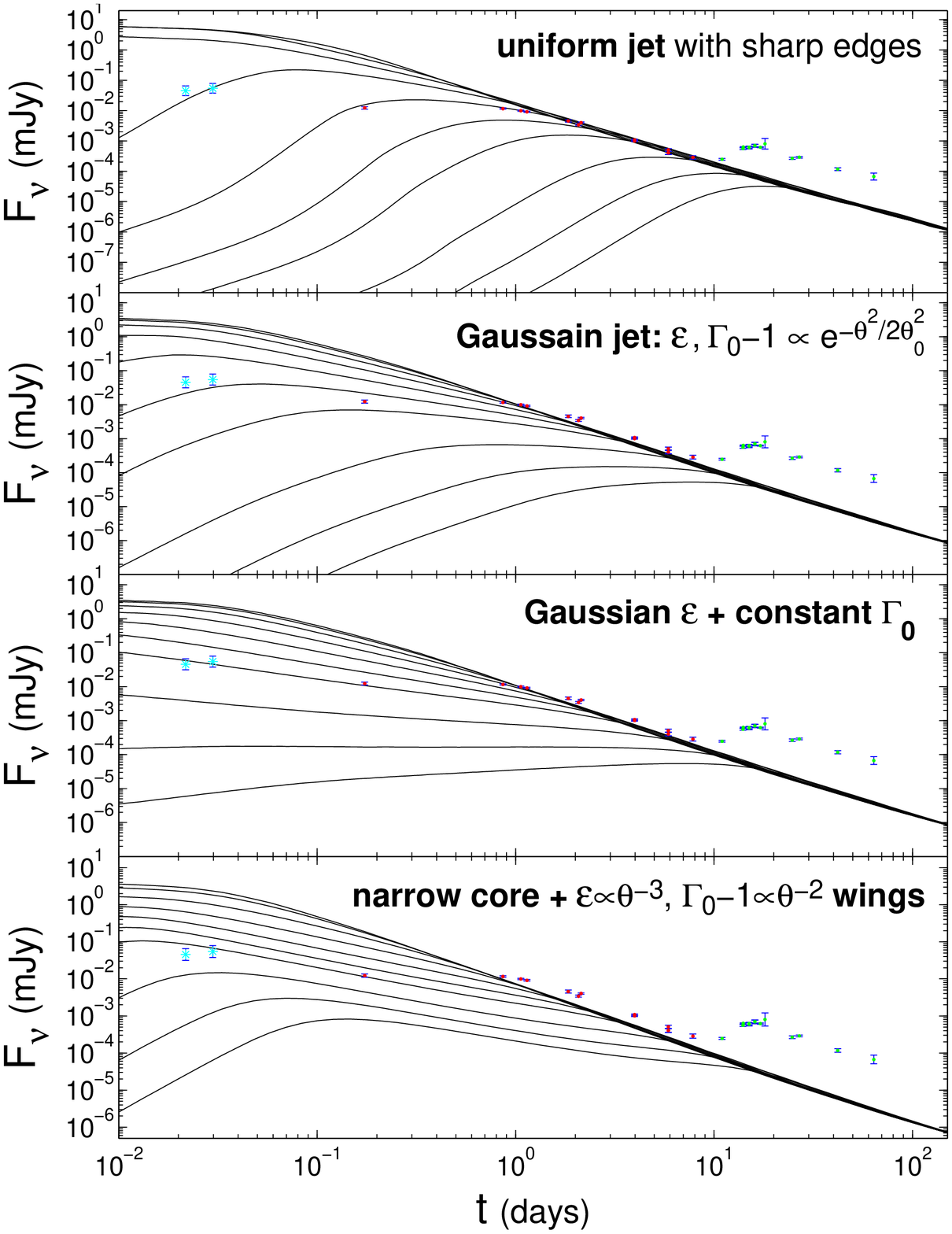} \caption{\label{XRF030723_4jets}The
R-band light curve of XRF 030723 is overlaid on top of theoretical
light curves for three different jet structures, and viewing angles
$\theta_{\rm
obs}/\theta_0=0,\,0.5,\,1,\,1.5,\,2,\,2.5,\,3,\,4,\,5,\,6$.  The
physical parameters are the same as in Fig. \ref{fit XRF030723}, where
the half opening angle of the uniform jet, $\theta_0=2.9^\circ$, is
identified with the core angle in the two other jet structures.}
\end{figure}

In addition to a uniform jet with sharp edges, we also consider other
jet structures: (i) a narrow core with power law wings
$\epsilon\approx\epsilon_0\min[1,(\theta/\theta_0)^{-3}]$,
$\Gamma_0-1\approx 299\propto\min[1,(\theta/\theta_0)^{-2}]$, and (ii)
a Gaussian jet with $\epsilon\propto\exp(-\theta^2/2\theta_0^2)$ and
either a constant $\Gamma_0=200$ or a Gaussian $\Gamma_0-1$ [i.e.
$\Gamma_0=1+199\exp(-\theta^2/2\theta_0^2)$].  The light curves for
different viewing angles are shown in Fig.  \ref{XRF030723_4jets}. For
a jet with power law wings where $\epsilon\propto\theta^{-a}$ it is
hard to reproduce the very flat light curve at early times that is
observed in XRF 070323 (and in XRGRB 041006), even for $a\approx 3$
that is expected for the collapsar model \citep{ZWH04}, and a steeper
drop in $\epsilon$ (i.e. a larger value of $a$) is required. For a
Gaussian jet, the light curves have a stronger dependence on the
angular profile of the initial Lorentz factor, $\Gamma_0(\theta)$.  If
it is constant, then the deceleration time at large viewing angles is
still very small, and the contribution to the observed flux from
material along the line of sight dominates at early times out to
reasonably large viewing angle $\theta_{\rm obs}/\theta_0\sim{\rm a\
few}$. If, on the other hand, it has a Gaussian profile,
$\Gamma_0-1\propto\exp(-\theta^2/2\theta_0^2)$, then the deceleration
time at large angles becomes large and the observed flux is dominated
by emission from the jet core even at early times.  This causes a rise
in the observed flux at early times for viewing angles outside the
core of the jet, similarly to a uniform jet viewed off-axis
\citep{KG03}, and in better agreement with the initially very flat
light curves of XRF 070323 and XRGRB 041006. We consider a Gaussian
profile for the kinetic energy per unit mass, $\Gamma_0-1$, to be more
realistic than a constant $\Gamma_0$, since the latter requires a
Gaussian profile for the rest mass per unit solid angle, $\mu$, that
is entrained in the outflow [since $\epsilon=(\Gamma_0-1)\mu c^2$],
while the former implies a constant $\mu$. If anything, one might
expect $\mu$ to increase with $\theta$ rather than decrease with
$\theta$ (since a larger amount of mass in the ejecta might be
expected near the walls of the funnel). Thus, from the models we
considered, a reasonable fit to the light curve of XRF 030723 (and
XRGRB 041006) can be obtained either for a uniform jet with sharp
edges viewed off-axis or for a Gaussian jet with a Gaussian profile in
both $\epsilon$ and $\Gamma_0-1$ viewed from outside its core.

The fact that the afterglow light curve of an XRGRB requires a viewing
angle that is only slightly outside the edge of the jet, while the
afterglow light curve of an XRF requires a larger viewing angle
($\theta_{\rm obs}\sim 2\theta_0$) provides a consistent picture where
a roughly uniform jet with relatively sharp edges is viewed as a GRB
from within the jet aperture [i.e. $\gamma(\theta_{\rm
obs}-\theta_0)\lesssim 1$], as an XRGRB from slightly outside the edge
of the jet [i.e. $1\lesssim\gamma(\theta_{\rm
obs}-\theta_0)\lesssim{\rm a\ few}$], and as an XRF from yet larger
off-axis viewing angles [i.e. $\gamma(\theta_{\rm
obs}-\theta_0)\gtrsim{\rm a\ few}$].

\subsection{Other events with sparse data}
\label{other_obs}

Besides the two events discussed above (in \S \ref{XRGRB041006} and \S
\ref{XRF030723}) there have been a few other XRFs with candidate
afterglow detections. The data in these case are, however, too sparse
to allow any meaningful constraint on theoretical models.

XRF 020903, detected by {\it HETE-II}, had an exceptionally low peak
energy of $\sim 5$ KeV. Detection of the optical and radio afterglow
was reported by \citet{Soderberg04}, together with the identification
of the likely host galaxy at $z=0.251$. Due to the large error box,
and the proximity to two other transient sources (which delayed prompt
identification), the optical light curve at early times was not well
sampled. The first detection is at $t=0.9\;$days after the burst,
while later observations are dominated by the light from the host.  In
contrast to the sparse optical measurements, the radio light curve was
extensively monitored with VLA over the period 25-370 days.  The
source, which was monitored at frequencies of 1.5, 4.9, 8.5 and
$22.5\;$GHz, was found to have a temporal index $\alpha$ similar to
that of ``standard'' GRBs \citep{Frail03}.

XRF 020427 was detected by {\it BeppoSAX}, and no redshift measurement
is available. There is a detection of X-ray emission at $t<100\;$s and
a later detection at $t\sim 1\;$day. If the last of the early time
detections (at $t\sim 50\;$s) is indeed marking the begin of the
afterglow \citep[as suggested by][]{Amati04}, then the inferred steep
afterglow decline would be hard to reconcile with a sharp edge seen
off-axis. However, given the lack of coverage, it is not clear whether
the detection at $t\sim 50\;$s is indeed part of the afterglow or,
instead, still a component of the prompt emission. In the latter
situation, with only one X-ray detection available, there is not much
that can be said in terms of possible models.

Other cases with possible counterparts are XRF 040912, which has a
candidate X-ray afterglow between $13.57\;$hr and $38.65\;$hr, and XRF
040916, which has an optical afterglow candidate but with no X-ray
detection.

\subsection{Supernova signatures in XRFs}
\label{SN}

The combined results on SN1998bw and SN2003dh offer the most direct
evidence yet that typical, long-duration, energetic GRBs result from
the deaths of massive stars \citep[e.g.,][]{Hjorth03,Stanek03}.  The
lack of hydrogen lines in both spectra is consistent with model
expectations that the star lost its hydrogen envelope to become a
Wolf-Rayet star before exploding. The broad lines are also suggestive
of an asymmetric explosion viewed along the axis of most rapid
expansion \citep{Mazzali01,ZWH04}. Despite the rather large
uncertainty on the true event rate of GRBs, a comparison with the
event rate of Type Ib/c SNe suggests that only a small
fraction,\footnote{The estimates range from $f_{\rm GRB}\sim 10^{-5}$
for the universal structured jet model, to $f_{\rm GRB}\approx (0.6\pm
0.2)\times 10^{-3}$ for the uniform jet model \citep{GR-R04}, where
the latter is the relevant one for the off-axis jet model.} $f_{\rm
GRB}\lesssim 10^{-3}$, of such SNe produce GRBs. The Type Ic SNe that
are firmly associated with GRBs are very bright Type Ic events, with
SN~1998bw being the brightest. The lack, however, of a SN in GRBs
010921 \citep{p03} and 020410 \citep{levan04a} to a limit of $\sim$1.5
and $\sim 2$ magnitudes fainter than SN 1998bw, respectively, suggests
that we may be seeing a broader luminosity function for the Type Ic
SNe that are associated with GRBs.

If the unification hypothesis discussed here is true (or in any model
where GRBs and XRFs are intrinsically the same object), XRFs should be
accompanied by a SN\citep{ZWH04} brightening in their afterglow light
curves, as seen in GRBs. Unfortunately, the sample of XRFs with known
redshifts and optical afterglows that are sufficiently well monitored
is very limited, with one possible exception -- XRF~030723 which has a
well sampled light curve but no measured redshift. There are,
nevertheless, both upper and lower limits on the redshift of XRF
030723. A lower limit of $z \gtrsim 0.3$ has been derived from the
non-detection of its host galaxy \citep{Fynbo04a}, while an upper
limit of $z< 2.3$ was derived from the lack of Ly$\alpha$ absorption.
\citet{Fynbo04a} obtained optical photometry and spectroscopy of XRF
030723, and found that the optical counterpart showed a ``bump" in the
light curve which may be the signature of a SN component. As discussed
in \S \ref{XRF030723}, the temporal and spectral energy distribution
evolution are hard to reconcile with other interpretations such as a
refreshed shock or a density variation in the external medium. For the
redshift range $z \sim 0.3-1$, all possible SN models require a rather
small mass of synthesized $^{56}$Ni \citep{Tominaga04}. This is
because the SN brightness at this distances is $\sim 2$ magnitudes
fainter than SN 1998bw. As the SN peak luminosity scales roughly
linearly with its $^{56}$Ni yield, we would expect very little
$^{56}$Ni production from a very faint SN.

The SN associated with XRF 030723 therefore appears to have properties
similar to those associated with GRB 010921 and GRB 020410, i.e. it
seems to lie at the low end of the hypernova luminosity function, and
is perhaps even closer in its properties to a normal Type Ic SN. This
might potentially be caused by our off-axis viewing angle which
resulted not only in an XRF instead of a GRB, but also in a dimmer SN
as opacity effects prevented us from seeing the brightest part of the
SN ejecta which lies along the rotational axis.  \citet{Nomoto03} find
that for the SNe that are associated with GRBs (or hypernovae), a
significant decrease in luminosity may occur only for viewing angles
$\theta_{\rm obs}\gtrsim 30^\circ$. This is a direct consequence of
the anisotropic distribution of the SN ejecta.  We find, however, that
for XRF 030723 $\theta_{\rm obs}\sim 2\theta_0\sim 6^\circ$, which is
well below $30^\circ$. Thus, the SN associated with XRF 030723 is
probably intrinsically dimmer than SN 1998bw. Clearly, more data on
the SN-GRB/XRF connection are necessary before we can understand the
full extent of the relation between these phenomena.

There is already some tentative evidence that a number of XRFs (011030
and 020427), for which no optical afterglow was detected, also have no
evidence for an associated SN \citep{levan04b}.  SNe such as SN~1998bw
would have been visible out to $z \sim 1.5$ in each case, while
somewhat fainter SNe would have been visible to $z \sim 1$.  Although
it is possible that these XRFs lie at $z\gtrsim 1$, it is still
puzzling given our attempt to tentatively identify GRBs, XRFs, and SNe
as similar objects observed with small, medium, and large inclination,
respectively.

A possibility which can explain both a relatively low redshift and the
absence of a SN detection is that the afterglows were heavily dust
extinguished.  In the off-axis jet model, prompt and intense X-ray/UV
radiation from the reverse shock may efficiently destroy and clear the
dust \citep{WD00,FKR01,PL02} in the circumburst cloud within the solid
angle corresponding to the initial jet aperture, i.e. at
$\theta<\theta_0$.  This implies relatively little extinction for
on-axis viewing angles, $\theta_{\rm obs}<\theta_0$ (practically no
extinction of emission from $\theta<\theta_0$ and a gradual increase
in the extinction as $\theta$ increases above $\theta_0$) but a
relatively large extinction for off-axis viewing angles, $\theta_{\rm
obs}>\theta_0$, especially for emission arising from
$\theta>\theta_0$.  Interestingly enough, in this case, there could be
many more obscured XRF optical afterglows, compared to GRB optical
afterglows.

\section{Conclusions}
\label{conc}

The existing XRF models have been examined and their predictions
tested against the afterglow observations of XRF 030723 and XRGRB
041006, the events with the best monitored afterglow light curves to
date within their respective class. We find that most models failed to
reproduce the very flat part observed in their early afterglow light
curve. This behavior is, however, naturally produced by a uniform jet
viewed off-axis (i.e. from $\theta_{\rm obs}>\theta_0$). The edge of
the jet must be sufficiently sharp, so that the emission at early
times would be dominated by the core of the jet, rather than by
material along the line of sight. Even for a jet with a narrow core
and wings where the energy per solid angle drops as
$\epsilon\propto\theta^{-3}$, as expected in the collapsar model, the
afterglow light curves at early times are not quite as flat as those
observed in XRF 030723 and XRGRB 041006. A Gaussian jet can produce a
sufficiently flat light curve at early times as long as both
$\epsilon$ and $\Gamma_0-1$ have a Gaussian profile (but not for a
constant initial Lorentz factor $\Gamma_0$; see Fig.
\ref{XRF030723_4jets}).

The afterglow light curve of XRGRB 041006 requires $(\theta_{\rm
  obs}-\theta_0)\sim 0.15\theta_0\sim 0.8\times 10^{-2}\;$rad, while
that of XRF 030723 requires $(\theta_{\rm obs}-\theta_0)\sim\theta_0
\sim 3^\circ\sim 0.05\;$rad. This supports a unified picture for GRBs,
XRGRBs and XRFs, where they all arise from the same narrow and roughly
uniform relativistic jets with reasonably sharp edges, and differ only
by the viewing angle from which they are observed. Within this scheme,
GRBs, XRGRBs and XRFs correspond to $\gamma(\theta_{\rm
  obs}-\theta_0)\lesssim 1$, $1\lesssim\gamma(\theta_{\rm
  obs}-\theta_0)\lesssim{\rm a\ few}$, and $\gamma(\theta_{\rm
  obs}-\theta_0)\gtrsim{\rm a\ few}$, respectively.
  
The empirical classification scheme by which an event is tagged as a
GRB, XRGRB or XRF (see \S \ref{clas}) is rather arbitrary. Therefore
there could be some cases where a jet that is viewed on-axis
($\theta_{\rm obs}<\theta_0$) will be classified as an XRGRB or XRF
instead of as a GRB, or the opposite case in which a jet viewed
off-axis ($\theta_{\rm obs}>\theta_0$) might be classified as a GRB
instead of as an XRGRB or an XRF. A more physically motivated
classification would be according to the ratio of the viewing angle
$\theta_{\rm obs}$ and the jet half-opening angle $\theta_0$ [e.g.,
on-axis events versus off-axis events, where off-axis events could
further be classified according to the value of $\gamma(\theta_{\rm
obs}-\theta_0)$], instead of relying purely on spectral
characteristics as in the present empirical scheme.  Such a
classification would, however, be much harder to implement as it is
not a trivial task to accurately determine the viewing angle.
  
Future observations with {\it HETE-II} and the recently launched {\it
Swift} satellite, will allow us to further test this picture, and
might also provide us with the necessary information to test the
structure of the jet. The strongest constraints could be obtained from
afterglow light curves of XRFs and XRGRBs that are well monitored from
early times and at various frequencies (ranging from radio to X-rays).
A useful complimentary method for constraining the jet structure is
via the statistics of the observed jet break times $t_j$ in the
afterglow light curves and the corresponding viewing angle
$\theta_{\rm obs}$ in the universal structured jet model or the jet
half-opening angle $\theta_0$ in the uniform jet model
\citep{PSF03,NGG04,LWD04}.

Similarly, the large statistical sample of GRBs and XRFs with redshift
that will be available during the {\it HETE-II/Swift} era, will allow
a reconstruction of the intrinsic luminosity function of the prompt
emission.  If GRBs, XRGRBs and XRFs are only a manifestation of the
viewing angle for a structured, universal jet (whose wings are
producing the XRFs), then no break would be expected in the luminosity
function. On the other hand, if GRBs are the results of viewing angles
that intersect the jet (whether structured or not), while XRFs and
XRGRBs are off-axis events, then one would naturally expect a break in
the luminosity function.  \citet{Guetta04} found that a luminosity
function with a break is favored in order for the predicted rate of
local bursts to be consistent with the observed rate. This also
prevents the existence of an exceedingly large number of GRB remnants
in the local Universe \citep{LP98, PRL00}.

The relative fraction of XRFs and XRGRBs to GRBs is also expected to
be different in the various models \citep{Lamb05}. If indeed an XRF
corresponds to $\gamma(\theta_{\rm obs}-\theta_0)\sim{\rm a\ few}$ and
$(\theta_{\rm obs}-\theta_0)\lesssim\theta_0$, the the solid angle
from which an XRF is seen scales as $\theta_0/\gamma$ or as $\theta_0$
for a constant $\gamma$ (at a constant distance to the source), while
the solid angle from which a GRB is seen scales as $\theta_0^2$.
Therefore, the ratio of solid angles for GRBs and XRFs scales as
$\theta_0$, and more GRBs compared to XRFs would be seen for larger
$\theta_0$. As the distance to the source increases, XRFs could be
detected only out to a smaller off-axis viewing angle, while most GRBs
would still be bright enough to be detected out to reasonably large
redshifts. Therefore, the ratio of GRBs to XRFs should increase with
redshift. Finally, if the true energy $E$ in the jet is roughly
constant, then the maximal redshift out to which a GRB could be
detected would decrease with $\theta_0$ since
$E_{\rm\gamma,iso}\propto\theta_0^{-2}$. This would increase the
statistical weight of narrow jets in an observed sample, as they could
be seen out to a larger volume.

We now briefly mention a few possible implication of the off-axis
model for XRFs and XRGRBs.  For sufficiently large viewing angles
outside the edge of the jet, one might expect some decrease in the the
variability of the prompt emission.  This is since the width of an
individual spike in the light curve scales as $\Delta
t\propto\delta\sim[\gamma(\theta_{\rm obs}-\theta_0)]^2$ while the
peak photon energy and fluence scale as $E_p\propto\delta^{-1}$ and
$f\propto\delta^{-3}$, respectively. Since the interval between
neighboring spikes in the light curve is typically comparable to the
width of an individual spike, $\Delta t$, then if $\Delta t$ increases
significantly for large viewing angles this would cause at least some
overlap between different pulses which would smear out some of the
variability. Thus one might expect XRFs to be somewhat less variable
than GRBs, at least on average, where a lower variability might be
expected for lower values of $E_p$. This may lead to a simple physical
interpretation of the observed variability-luminosity relation in the
prompt gamma-ray/X-ray emission \citep{FR00,Reic01}.

Another possible signature of the off-axis model for XRFs is in the
reverse shock emission.  If the reverse shock is at least mildly
relativistic, then the optical flash emission would be less beamed
than the prompt X-ray or gamma-ray emission, due to the deceleration
of the ejecta by the passage of the reverse shock. This might cause
the optical flash to be suppressed by a smaller factor relative to the
gamma-ray emission, compared to the corresponding on-axis fluxes. Thus
XRFs or XRGRBs might still show reasonably bright optical emission
from the reverse shock, which might in some cases be almost as bright
as for classical GRBs. Finally, XRFs and XRGRBs might also show a
larger degree of polarization compared to GRBs (see \S \ref{pol}).

\begin{figure}
\plotone{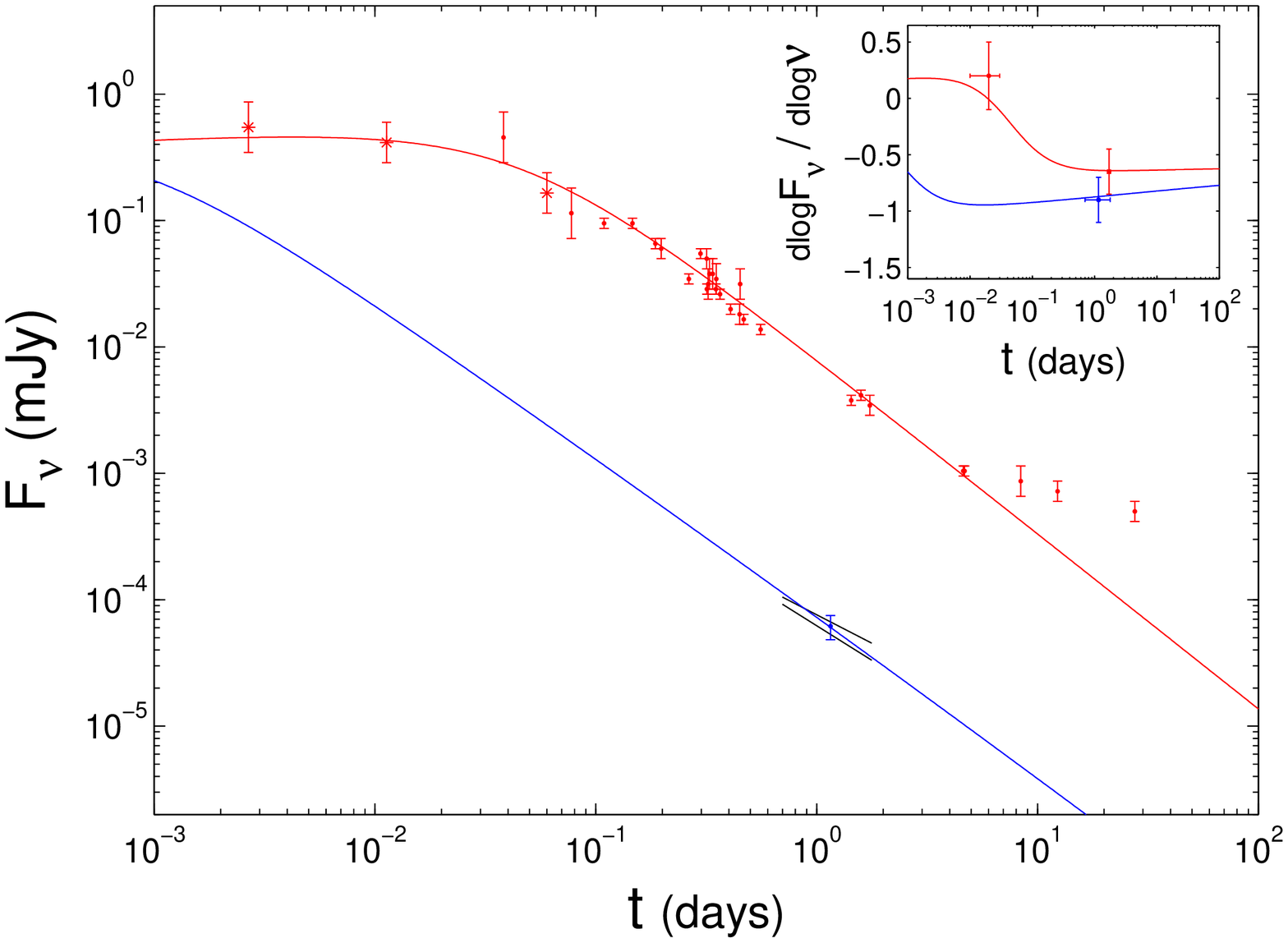}
\caption{\label{XRGRB041006_sph} A tentative fit to the afterglow
  observations of XRGRB 041006, presented in the same format as Fig.
  \ref{fit_XRGRB041006}, but using a different theoretical model.  The
  model used here is taken from 
\citet{GS02}, and features a spherical
  afterglow shock going into a stellar wind external medium ($k=2$),
  with $p=2.2$. It also provides an adequate description of a jet
  (either structured or uniform) before the jet break time (i.e. at
  $t<t_j$).  The remaining four model parameters ($E_{\rm iso}$,
  $A_*$, $\epsilon_e$, and $\epsilon_B$) cannot be uniquely
  determined, as there are effectively only three constraints (the
  flux normalization and location of two break frequencies:
  $\nu_m$ and $\nu_c$).
}
\end{figure}

An important conclusion from this study is that jet models in which
$\epsilon$ and $\Gamma_0$ vary smoothly inside the jet, and where our
lines of sight are within the jet, do not naturally reproduce the
afterglow light curves of XRF 030723 and XRGRB 041006.  The best
example of such a model is the ``universal structured jet'' model
\citep{LPP01,RLR02,ZM02b}, where both $\epsilon$ and $\Gamma_0$ vary
smoothly as a power law in $\theta$ (the usual assumption being that
$\epsilon\propto\theta^{-2}$ outside of some core angle). This model
fails to account for the very flat initial part of the afterglow light
curves and its subsequent decay. 

A possible way around this problem might be to identify the flat part
of the light curve with the passage of the break frequency $\nu_m$
through the optical band.  This should be accompanied by a change in
the optical spectral slope, and should not be observed in other
frequency ranges such as the radio or X-rays. For XRGRB 041006 this
may actually provide a viable explanation for the data (see Fig.
\ref{XRGRB041006_sph}).  For XRF 030723, however, a similar model
fails because it does not reproduce both of the observed values of the
temporal index $\alpha_{\rm op}$ (before or after the passage of
$\nu_m$) or the observed spectral slope $\beta_{\rm op}$. One could in
principle invoke both a jet break and the passage of a break frequency
at roughly the same time, for a jet viewed on-axis.  This would
require that $\nu_m\sim\nu_c\sim\nu_{\rm op}$ at $t_0\sim t_j\sim
0.1-1\;$days, which is a large coincidence and is therefore
unlikely.\footnote{Such constraints would not leave enough free model
parameters in order to also account for the temporal decay index in
the X-rays, $\alpha_X$.}  Even if this was the case, this assumption
would be hard to reconcile with the measured optical spectral slope of
$\beta_{\rm op}=0.96\pm 0.04$ at $t=1.13\;$days, as the spectral break
frequencies would still be near the optical at that time, resulting in
a smaller value of $\beta_{\rm op}$.  The afterglow light curve of XRF
030723 therefore provides evidence against this class of models.

\acknowledgments

We thank N. Butler, C. Kouveliotou, S. Woosley for useful
discussions. This research was supported by US Department of Energy
under contract number DE-AC03-76SF00515 (J.G.) and by NASA through a
Chandra Postdoctoral Fellowship award PF3-40028 (E. R.-R.). Part of
this work was done while E. R.-R. and J.G. were visiting the UCSC.


\begin{thebibliography}{99}


\bibitem[Amati et al.(2002)]{Amati02} Amati, L.  et al. 2002, A\&A,
390, 81
 
\bibitem[Amati et al.(2004)]{Amati04} Amati, L.  et al. 2004, A\&A
submitted (astro-ph/0407166)

 
\bibitem[Ayani et al.(2004)]{Ayani04} Ayani, K., et al. 2004, GCN,
2779

\bibitem[Band et al. (1993)]{Band93} Band, D.~L., et al. 1993, ApJ,
413, 281

\bibitem[Band \& Preece(2005)]{Band05} Band, D.~L., \& Preece
R.~D. 2005, submitted to ApJ  (astro-ph/0501559)

\bibitem[Barnard et al.(2004a)]{Barnard04a} Barnard, V., et
al. 2004a, GCN, 2774 

\bibitem[Barnard et al.(2004b)]{Barnard04b} Barnard, V., et al. 2004b,
GCN, 2786

\bibitem[Barraud et al.(2003)]{Barraud03} 
Barraud, C. et al. 2003, A\&A, 400, 1021

\bibitem[Bikmaev et al.(2004)]{Bikmaev04} Bikmaev, I., et al. 2004,
GCN, 2826

\bibitem[Blain \& Natarajan (2000)]{BN00}
Blain, A. W., \& Natarajan,
P. 2000, MNRAS, 312, L35

\bibitem[Bromm \& Loeb (2002)]{BL02}
Bromm, V., \& Loeb, A. 2002, ApJ,
575, 111

\bibitem[Butler et al.(2004a)]{Butler04a} Butler, N., Dullinghan, A.,
Ford, P., Ricker, G., Vanderspek, R., Hurley, K., Jernigan, J., Lamb,
D., Graziani, C. 2004a, ApJ in press (astro-ph/0401020)

\bibitem[Butler et al.(2004b)]{Butler04b}
Butler, N., et al. 2004b, GCN, 2808

\bibitem[Covino et al.(2004)]{Covino04} Covino, S., et al. 2004, GCN,
2803

\bibitem[Da Costa \& Noel(2004)]{DaCosta04} Da Costa, P., \& Noel,
N. 2004, GCN, 2789

\bibitem[Dado, Dar, \& De R\'ujula(2004)]{DDD04}
Dado, S., Dar, A. \& De R\'ujula, A. 2004, A\&A, 422, 381

\bibitem[Dalal, Griest \& Pruet(2002)]{DGP02}
Dalal, N., Griest, K., \& Pruet, J. 2002, ApJ, 564, 209

\bibitem[D'Avanzo et al.(2004)]{D'Avanzo04} D'Avanzo, P., et
al. 2004, GCN, 2788

\bibitem[Dermer, Chiang \& B\"ottcher(1999)]{DCB99}
Dermer, C.~D., Chiang J., \& B\"ottcher, M. 1999, ApJ, 513, 656

\bibitem[Drenkhahn(2002)]{Drenkhahn02}
Drenkhahn, G. 2002, A\&A, 387, 714

\bibitem[Eichler \& Levinson(2004)]{EL04}
Eichler, D., \& Levinson, A. 2004, ApJ, 614, L13

\bibitem[Fan, Wei \& Wang(2004)]{FWW04}
Fan, Y. Z., Wei, D. M., \& Wang, C. F. 2004, MNRAS, 351, L78

\bibitem[Fenimore \& Ramirez-Ruiz (2000)]{FR00} Fenimore, E. E., \&
Ramirez-Ruiz, E. 2000, (astro-ph/0004176)

\bibitem[Ferrero et al.(2004)]{Ferrero04} Ferrero, P., et al. 2004,
GCN, 2777

\bibitem[Fox et al.(2003)]{Fox03}
Fox, D. B., et al. 2003, GCN, 2343

\bibitem[Frail et al.(2003)]{Frail03}
Frail, D. A., Kulkarni, S. R.,
Berger, E. \& Wieringa, M. H. 2003, AJ, 125, 2299

\bibitem[Fruchter, Krolik \& Rhoads(2001)]{FKR01}
Fruchter, A., Krolik, J. H., Rhoads, J. E. 2001, ApJ, 563, 597

\bibitem[Fugazza et al.(2004)]{Fugazza04} Fugazza, D., et al. 2004,
GCN, 2782

\bibitem[Fukushi et al.(2004)]{Fukushi04} Fukushi, H., et al. 2004,
GCN, 2767

\bibitem[Fynbo et al.(2004a)]{Fynbo04a} Fynbo, J. P. U. et al. 2004a,
preprint (astro-ph/0402240)

\bibitem[Fynbo et al.(2004b)]{Fynbo04b} Fynbo, J.~P.~U. et al. 2004b,
GCN, 2802

\bibitem[Galassi et al.(2004)]{Galassi04}
Galassi, M., et al. 2004, GCN, 2770

\bibitem[Garg et al.(2004)]{Garg04} Garg, A., et al. 2004, GCN, 2829

\bibitem[Ghirlanda, Ghisellini \& Lazzati(2004)]{GGL04} Ghirlanda, G.,
Ghisellini, G., \& Lazzati, D. 2004, ApJ in press, astro-ph/0405602
 
\bibitem[Ghisellini \& Lazzati(1999)]{GL99}
Ghisellini, G., \& Lazzati, D. 1999, MNRAS, 309, L7 

\bibitem[Greco et al.(2004)]{Greco04} Greco, G., et al. 2004, GCN,
2804

\bibitem[Granot(2003)]{Granot03}
Granot, J. 2003, ApJ, 596, L17

\bibitem[Granot \& Loeb(2003)]{GL03}
Granot, J., \& Loeb, A. 2003, ApJ, 593, L81

\bibitem[Granot \& K\"onigl(2003)]{GKo03}
Granot, J., \& K\'onigl, A. 2003, ApJ, 594, L83

\bibitem[Granot \& Kumar(2003)]{GK03}
Granot, J., Kumar, P. 2003, ApJ, 591, 1086

\bibitem[Granot et al.(2001)]{Granot01}
Granot, J, Miller, M., Piran, T., Suen, W.~M., \& Hughes, P.~A.
2001, in Gamma-Ray Bursts in the Afterglow Era, ed. E. Costa, F.
Frontera, \& J. Hjorth (Berlin: Springer), 312

\bibitem[Granot et al.(2002)]{Granot02}
Granot, J., Panaitescu, A., Kumar, P., Woosley, S. E. 2002, ApJ, 570, L61

\bibitem[Granot \& Sari(2002)]{GS02}
Granot, J., \& Sari, R. 2002, ApJ, 568, 820

\bibitem[Granot \& Taylor(2005)]{GT05}
Granot, J., \& Taylor, G. B. 2005, ApJ in press (astro-ph/0412309)

\bibitem[Granot \& Ramirez-Ruiz(2004)]{GR-R04}
Granot, J., \& Ramirez-Ruiz, E. 2004, ApJ, 609, L9

\bibitem[Gruzinov(1999)]{Gruzinov99}
Gruzinov, A. 1999, ApJ, 525, L29

\bibitem[Guetta, Granot \& Begelman(2005)]{GGB05}
Guetta, D., Granot, J., \& Begelman, M.~C. 2005, ApJ in press 
(astro-ph/0407063)

\bibitem[Guetta et al. (2004)]{Guetta04} 
Guetta, D., Perna, R., Stella, L. \& Vietri, M. 2004, ApJL, 615, 73 

\bibitem[Heise et al.(2001)]{Heise01}
Heise, J., in't Zand, J., Kippen, R. M., \& Woods, P. M. 2001, in Proc.
of the conference ``Gamma-ray Bursts in the Afterglow Era'', 16

\bibitem[Heyl \& Perna(2003)]{heyl} Heyl, J. S., \& Perna, R. 2003,
ApJ 586, L13

\bibitem[Hjorth et al. (2003)]{Hjorth03} Hjorth, J., et al. 2003,
Nature, 423, 847

\bibitem[Huang et al.(2002)]{Huang02}
Huang, Y. F., Dai, Z. G. \& Lu, T. 2002, MNRAS, 332, 735

\bibitem[Huang et al.(2004)]{Huang04}
Huang, Y. F., Wu, X. F., Dai, Z. G., Ma, H. T., Lu, T. 2004, ApJ,
605, 300

\bibitem[Kahharov et al.(2004)]{Kahharov04} Kahharov, B., et
al. 2004, GCN, 2775

\bibitem[Kelson et al.(2004)]{Kelson04}
Kelson, D.~D., Koviak, K., Berger, E., \& Fox, D.~B. 2004, GCN circ. 2627 

\bibitem[Kinoshita et al.(2004)]{Kinoshita04} Kinoshita, D., et
al. 2004, GCN, 2785

\bibitem[Kippen et al.(2003)]{Kippen03} 
Kippen, R.~M., et al. 2003, in Gamma-Ray Bursts ans Afterglow
Astronomy, AIP Conf. Proceedings 662, ed. G.~R. Ricker \& 
R.~K. Vanderspek (New York: AIP), 25

\bibitem[Klotz et al.(2004)]{Klotz04} Klotz, A., et al. 2004, GCN,
2784

\bibitem[Kouveliotou et al. (2004)]{CK04} Kouveliotou, C., et
al. 2004, ApJ, 608, 872

\bibitem[Kumar \& Piran(2000)]{kumar} Kumar, P., \& Piran, T. 2000,
ApJ 535, 152

\bibitem[Kumar \& Granot(2003)]{KG03}
Kumar, P., \& Granot, J.  2003, ApJ, 591, 1075

\bibitem[Lamb et al.(2004)]{Lamb04}
Lamd, D.~Q., et al. 2004, New Astron. Rev., 48, 423

\bibitem[Lamb et al.(2005)]{Lamb05}
Lamb, D., Q., Donaghy, T., Q. \& Graziani, C. 2005, ApJ, 620, 355

\bibitem[Lazzati et al.(2002)]{lazzati2} Lazzati, D., Rossi, E.,
  Covino, S., Ghisellini, G., \& Malesani, D. 2002, A\&A 396, L5

\bibitem[Levan et al. (2004a)]{levan04a} Levan, A. et. al. 2004, ApJ in
press (astro-ph/0403450)

\bibitem[Levan et al. (2004b)]{levan04b} Levan, A. et. al. 2004, ApJ in
press (astro-ph/0410560)

\bibitem[Levinson et al.(2002)]{Levinson02}
Levinson, A., Ofek, E.~O., Waxman, E., \& Gal-Yam, A. 
2002, ApJ, 576, 923

\bibitem[Liang, Wu \& Dai (2004)]{LWD04} Liang, E. W., Wu, X. F. \&
Dai, Z. G. 2004, MNRAS, 354, 81

\bibitem[Lipunov, Postnov \& Prokhorov(2001)]{LPP01}
Lipunov, V. M., Postnov, K. A., \& Prokhorov, M. E. 2001, Astron.
Rep., 45, 236

\bibitem[Lloyd-Ronning, Fryer, \& Ramirez-Ruiz(2002)]{LRFRR02}
Lloyd-Ronning, N., Fryer, C., \& Ramirez-Ruiz, E. 2002, ApJ, 574, 554

\bibitem[Lloyd-Ronning \& Ramirez-Ruiz(2002)]{L-RR-R02}
Lloyd-Ronning, N., \& Ramirez-Ruiz, E. 2002, ApJ, 576, 101

\bibitem[Loeb \& Perna (1998)]{LP98} 
Loeb, A. \& Perna, R. 1998, ApJL, 503, 35

\bibitem[Lyutikov, Pariev \& Blandford(2003)]{LPB03}
Lyutikov, M., Pariev, V. I., \& Blandford, R. D. 2003, ApJ, 597, 998


\bibitem[Mazzali et al. (2001)]{Mazzali01}Mazzali, P. A., Nomoto, K.,
Patat, F., \& Maeda, K. 2001, ApJ, 559, 1047

\bibitem[Medvedev \& Loeb(1999)]{ML99}
Medvedev, M.~V., \& Loeb, A. 1999, ApJ, 1999, 526, 697

\bibitem[M\'esz\'aros et al.(1998)]{mes98} M\'esz\'aros, P., Rees,
M. J., \& Wijers, R. 1998, ApJ 499, 301

\bibitem[M\'esz\'aros et al.(2002)]{Meszaros02}
M\'esz\'aros, P., Ramirez-Ruiz, E., Rees, M.~J., \& Zhang, B.
2002, ApJ, 578, 812

\bibitem[Misra \& Pandey(2004a)]{Misra04a} Misra, K., \& Pandey,
S.~B. 2004a, GCN, 2794

\bibitem[Misra \& Pandey(2004b)]{Misra04b} Misra, K., \& Pandey,
S.~B. 2004b, GCN, 2795

\bibitem[Mochkovitch et al.(2003)]{moch03}
Mochkovitch, R., Daigne, F., Barraud, C., \& Atteia, J. L. 2003, ASP Conference
Series (San Francisco ASP), in press (astro-phj/0303289)

\bibitem[Moderski, Sikora \& Bulik(2000)]{MSB00}
Moderski, R., Sikora, M., \& Bulik, T. 2000, ApJ, 529, 151

\bibitem[Monfardini et al.(2004)]{Monfardini04} Monfardini, A., et
al. 2004, GCN, 2790

\bibitem[Nakar, Granot \& Guetta(2004)]{NGG04} Nakar, E., Granot J. \&
Guetta, D. 2004, ApJ, 606, L37

\bibitem[Nakar et al.(2003)]{nakar} Nakar, E., Piran, T., \& Granot,
J. 2003, NewA 8, 495

\bibitem[Nakar \& Piran (2005)]{Nakar05} Nakar, E., \& Piran, T. 2005,
ApJ submitted (astro-ph/0412232) 

\bibitem[Nakar, Piran \& Granot(2003)]{NPG03} 
Nakar, E., Piran, T., \& Granot, J. 2003, ApJ, 579, 699

\bibitem[Nakar, Piran \& Waxman(2003)]{NPW03} 
Nakar, E., Piran, T., \& Waxman, E. 2003, JCAP, 10, 5

\bibitem[Nomoto et al. (2003)]{Nomoto03} 
Nomoto, K., Maeda, K., Mazzali, P. A., Umeda, H., Deng, J. \&
Iwamoto, K.  2003, to appear in ``Stellar Collapse'' (Astrophysics
and Space Science; Kluwer) ed. C.  L. Fryer


\bibitem[Panaitescu et al.(1998)]{pmr98} Panaitescu, A., M\'esz\'aros,
P., \& Rees, M. J. 1998, ApJ 503, 314

\bibitem[Panaitescu \& Kumar(2000)]{PK00} Panaitescu, A., \& Kumar,
P. 2000, ApJ, 543, 66

\bibitem[Peng, K\"onigl \& Granot(2005)]{PKG05}
Peng, F., K\'onigl, A., \& Granot, J. 2005 ,submitted to ApJ
(astro-ph/0410384)

\bibitem[Perna \& Lazzati(2002)]{PL02}
Perna, R. \& Lazzati, D. 2002, ApJ, 580, 261

\bibitem[Perna \& Loeb(1998)]{PL98}
Perna, R. \& Loeb, A. 1998, 509L, 85

\bibitem[Perna, Raymond \& Loeb (2000)]{PRL00} Perna, R.,
  Raymond, J. \& Loeb, A.  2000, ApJ, 533, 658

\bibitem[Perna, Sari \& Frail(2003)]{PSF03}
Perna, R., Sari, R. \& Frail, D. 2003, ApJ, 594, 379

\bibitem[Price et al. (2003)] {p03} Price, P. et al. 2003, ApJ, 584,
931

\bibitem[Price et al.(2004a)]{Price04a} Price, P.~A., et al. 2004a,
GCN, 2771

\bibitem[Price et al.(2004b)]{Price04b}
Price, P.~A., et al. 2004b, GCN, 2791

\bibitem[Prochasksa et al.(2004)]{Prochaska04}
Prochaska, J.~X., et al. 2004, preprint (astro-ph/0402085)

\bibitem[Ramirez-Ruiz et al.(2001a)]{enrico} Ramirez-Ruiz, E., Dray,
  L. M., Madau, P., \& Tout, C. A. 2001a, MNRAS 327, 829

\bibitem[Ramirez-Ruiz et al.(2001b)]{rmr01} Ramirez-Ruiz, E., Merloni
A., \& Rees M. J. 2001b, MNRAS 324, 1147

\bibitem[Ramirez-Ruiz, Celotti \& Rees(2002)]{RR02} Ramirez-Ruiz, E.,
Celotti, A., \& Rees, M. J. 2002, MNRAS, 337, 1349

\bibitem[Ramirez-Ruiz \& Lloyd-Ronning(2002)]{RRLR02} 
Ramirez-Ruiz, E., \& Lloyd-Ronning, N. M. 2002, New Astron., 7, 197

\bibitem[Ramirez-Ruiz \& Madau(2004)]{R-RM04} Ramirez-Ruiz, E., \&
Madau, E. 2004, ApJ, 608, L89

\bibitem[Ramirez-Ruiz et al.(2005)]{R-R05} Ramirez-Ruiz, E., Granot,
J., Kouveliotou, C., Woosley, S.~E., Patel, S.~K., \& Mazzali,
P.~A. 2005, submitted to ApJL (astro-ph/0412145)

\bibitem[Reichart et al. (2001)]{Reic01} Reichart, D., Lamb, D.,
Fenimore, E. E., Ramirez-Ruiz, E., Cline, T., \& Hurley, K. 2001, ApJ,
552, 57

\bibitem[Rees \& M\'{e}sz\'{a}ros(1998)]{rm98} Rees, M. J., \&
M\'{e}sz\'{a}ros, P. 1998, ApJ 496, L1


\bibitem[Rhoads(1997)]{Rhoads97}
Rhoads, J. E. 1997, ApJ, 487, L1

\bibitem[Rhoads(2003)]{Rhoads03}
Rhoads, J. E. 2003, ApJ, 591, 1097

\bibitem[Richardson et al. (2002)]{rich02} Richardson, D. Branch, D.,
Casebeer, D., Millard, J., Thomas, R.C., Baron, E. 2002, ApJ, 123, 745


\bibitem[Rossi, Lazzati \& Rees(2002)]{RLR02} Rossi, E., Lazzati,
D. \& Rees, M. J. 2002, MNRAS, 332, 945

\bibitem[Sakamoto et al.(2004)]{Sakamoto04}
Sakamoto, T., et al. 2004, submitted to ApJ (astro-ph/0409128)

\bibitem[Sari(1999)]{Sari99}
Sari, R. 1999, ApJ, 524, L43

\bibitem[Sari \& Esin(2001)]{SE01}
Sari, R., \& Esin, A. A. 2001, ApJ, 548, 787

\bibitem[Sari, Piran \& Narayan(1998)]{SPN98} 
Sari, R., Piran, T., \& Narayan, R.\ 1998, \apjl, 497, L17 

\bibitem[Smith et al.(2003)]{Smith03}
Smith, D. A., Akerlof, C. W., \& Quimby, R. 2003, GCN circ. 2338

\bibitem[Soderberg, Berger \& Frail(2003)]{SBF03}
Soderberg, A. M., Berger, E. \& Frail, D. A. 2003a, GCN circ. 2330

\bibitem[Soderberg et al.(2004)]{Soderberg04}
Soderberg, A. M., et al. 2004, ApJ, 606, 994

\bibitem[Soderberg \& Frail(2004)]{SF04} Soderberg, A. M., \& Frail,
D. 2004, GCN 2787

\bibitem[Stanek et al.(2003)]{Stanek03} Stanek, K.~Z., et al. 2003,
ApJ, 591, L17

\bibitem[Thomsen et al.(2004)]{Thomsen04}
Thomsen, B., et al. 2004, preprint (astro-ph/0403451)

\bibitem[Tominaga et al.(2004)]{Tominaga04}
Tominaga, N., et al. 2004, ApJ, 612, L105

\bibitem[Totani \& Panaitescu(2002)]{TP02}
Totani, T., \& Panaitescu, A. 2002, ApJ, 576, 120

\bibitem[Wang \& Loeb(2000)]{wl} Wang, X., \& Loeb, A. 2000,  ApJ
535, 788

\bibitem[Waxman(2003)]{Waxman03}
Waxman, E. 2003, Nature, 423, 388

\bibitem[Waxman \& Draine (2000)]{WD00} Waxman, E., \& Draine, B. T.
2000 , ApJ, 537, 796

\bibitem[Watson et al.(2004)]{Watson04}
Watson, D., et al. 2004, preprint (astro-ph/0401225)

\bibitem[Woods \& Loeb(1999)]{WL99}
Woods, E., \& Loeb, A. 1999, ApJ, 523, 187

\bibitem[Yamazaki, Ioka \& Nakamura(2002)]{YIN02}
Yamazaki, R., Ioka, K., \& Nakamura, T. 2002, ApJ, 571, L31

\bibitem[Yamazaki, Ioka \& Nakamura(2003)]{YIN03}
Yamazaki, R., Ioka, K., \& Nakamura, T. 2003, ApJ, 593, 941

\bibitem[Yamazaki, Ioka \& Nakamura(2004a)]{YIN04a}
Yamazaki, R., Ioka, K., \& Nakamura, T. 2004a, ApJ, 606, L33

\bibitem[Yamazaki, Ioka \& Nakamura(2004b)]{YIN04b}
Yamazaki, R., Ioka, K., \& Nakamura, T. 2004b, ApJ, 607, L103

\bibitem[Yost et al.(2004)]{Yost04} Yost, S., et al. 2004, GCN, 2776

\bibitem[Zhang et al.(2004)]{Zhang04}
Zhang, B., Dai, X., Lloyd-Ronning, N.~M., \& M\'esz\'aros, P. 2004,
ApJ, 601, L119

\bibitem[Zhang \& M\'esz\'aros(2002a)]{ZM02a}
Zhang, B. \& M\'esz\'aros, P. 2002a, ApJ, 581, 1236

\bibitem[Zhang \& M\'esz\'aros(2002b)]{ZM02b}
Zhang, B. \& M\'esz\'aros, P. 2002b, ApJ, 571, 876

\bibitem[Zhang, Woosley \& Heger(2004)]{ZWH04}
Zhang, W., Woosley, S.~E., \& Heger, A. 2004, ApJ, 608, 365

\bibitem[Zhang, Woosley \& MacFadyen(2003)]{ZWM03}
Zhang, W., Woosley, S.~E., \& MacFadyen, A.~I. 2004, ApJ, 608, 365


\end{thebibliography}
\end{document}